\pdfoutput=1
\documentclass{JINST}
\let\ifpdf\relax

\usepackage{lineno}
\usepackage{color}
\usepackage{url}

\usepackage{graphicx}
\usepackage[figuresright]{rotating}

\title{The Liquid Argon Purity Demonstrator}

\author{M.~Adamowski$^a$,
B.~Carls$^a$,
E.~Dvorak$^b$,
A.~Hahn$^a$,
W.~Jaskierny$^a$,
C.~Johnson$^c$,
H.~Jostlein$^a$,
C.~Kendziora$^a$,
S.~Lockwitz$^a$,
B.~Pahlka$^a$, 
R.~Plunkett$^a$,
S.~Pordes$^a$,
B.~Rebel$^a$\thanks{Corresponding author: brebel@fnal.gov (B.~Rebel)},
R.~Schmitt$^a$,
M.~Stancari$^a$,
T.~Tope$^a$\thanks{Corresponding author: tope@fnal.gov (T.~Tope)},
E.~Voirin$^a$, and
T.~Yang$^a$ \\
\llap{$^a$}Fermi National Accelerator Laboratory, P.O. Box 500, Batavia, IL, 60510, USA \\
\llap{$^b$}South Dakota School of Mines \& Technology, 501 East Saint Joseph Street, Rapid City, SD 57701, USA \\
\llap{$^c$}Indiana University, 727 E. Third St., Swain Hall West, Room 117, Bloomington, IN 47405, USA }

\abstract{The Liquid Argon Purity Demonstrator was an R\&D test stand designed to determine if electron drift lifetimes adequate for large neutrino detectors could be achieved without first evacuating the cryostat.  We describe here the cryogenic system, its operations, and the apparatus used to determine the contaminant levels in the argon and to measure the electron drift lifetime.  The liquid purity obtained by this system was facilitated by a gaseous argon purge. Additionally, gaseous impurities from the ullage were prevented from entering the liquid at the gas-liquid interface by condensing the gas and filtering the resulting liquid before returning to the cryostat.  The measured electron drift lifetime in this test was greater than 6 ms, sustained over several periods of many weeks.  Measurements of the temperature profile in the argon, to assess convective flow and boiling, were also made and are compared to simulation. }

\begin{document}

\tableofcontents

\section{Introduction}

Liquid argon (LAr) time projection chambers (TPCs) provide a robust and elegant method for measuring the properties of neutrino interactions above a few tens of MeV by providing 3D event imaging with excellent spatial resolution.  The ionization electrons created by the passage of charged particles through the liquid are transported with typical diffusion of less than a millimeter per meter of drift distance by a uniform electric field over macroscopic distances.  Imaging is achieved by sets of parallel wires oriented in different directions and perpendicular to the drift field.  The signals induced by the drifting electrons on the wires are amplified and digitized by wave-form recording electronics.  The projection of a particle track in the plane perpendicular to the drift field, i.e. the plane of the wires, is given by the pattern of hits on the wire planes while the projection of the track in the plane parallel to the wires is given by the arrival time of the signals on the wires~\cite{Marx:1978zz,Gatti:1978bp}.  This technology requires that electrons drift without attachment to electronegative contaminants.  The time that a cloud of ionization electrons can drift before the fraction $1/e$ of the total ionization electrons attach to electronegative components is called the electron drift lifetime.  LArTPC technology has experienced renewed and strengthened interest since having recently been chosen as the preferred technology for the LBNE future long-baseline neutrino oscillation experiment~\cite{Akiri:2011dv}.  We begin with a brief overview of LArTPC development. 

The ICARUS Collaboration led a pioneering effort in the development of LArTPC technology culminating in the construction of the T600 LArTPC~\cite{ref:icarusFirstResults}.  The T600 LArTPC is housed in a 760 ton capacity cryostat that is surrounded by insulating layers of Nomex honeycomb cells~\cite{Amerio:2004ze}.  The cryostat was evacuated to a pressure of $10^{-4}$ mbar before filling with liquid argon.  Electron lifetimes greater than 6 ms were obtained with a contamination less than 0.050 parts per billion (ppb) oxygen equivalent.

The Materials Test Stand (MTS) at the Fermi National Accelerator Laboratory (Fermilab) was developed to evaluate the effect of different materials on electron lifetime~\cite{Andrews:2009zza}.  In this system a 250~L vacuum-insulated vessel was evacuated to a pressure of $10^{-6}$ Torr before it was filled.  The system employed commercial filter materials in a Fermilab-designed filter system to reduce contamination from water and oxygen and measured electron lifetimes of approximately 8 ms using a dedicated purity monitor.

The ArgoNeuT experiment at Fermilab was the first LArTPC in the United States to be placed in a neutrino beam~\cite{ref:argoneut}.  Commissioned in 2009, it had an 550 L vacuum insulated cryostat that was evacuated before filling with liquid argon.  The purification system purified argon gas boiled off in the gaseous region of the cryostat.  With this system, ArgoNeuT was able to obtain lifetimes of about 750 $\mu$s.

The ARGONTUBE LArTPC at The University of Bern was developed to investigate the ability to drift electrons over distances of up to 5 m \cite{ref:argontube}.  It uses a vacuum insulated cryostat and is evacuated to $5\times10^{-5}$ mbar before filling with liquid argon.  ARGONTUBE has been able to reach contamination levels down to 0.15 ppb and achieved lifetimes of 2 ms with a 240 V/cm drift field. 

The conventional liquid argon vessels described in this section were evacuated to remove water, oxygen, and nitrogen contaminants present in the ambient atmosphere prior to filling with liquid argon.  Physics requirements for long-baseline neutrino experiments dictate larger cryogenic vessels to hold bigger detectors; the mechanical strength required to resist the external pressure of evacuation becomes prohibitively costly for such large vessels.  Thus, the concept of purification without evacuation and testing with the Liquid Argon Purity Demonstrator (LAPD) was proposed in 2006~\cite{schmitt}.

The LAPD located at Fermilab was designed to achieve the ultra high purity required by LArTPCs in a vessel that cannot be evacuated.  The system relies heavily on the experience from the MTS~\cite{Andrews:2009zza} in its design and operation plan.  Previous studies suggest that the concentration of oxygen in a vessel purged with gaseous argon can be reduced to 100 ppm after 2.6 volume exchanges~\cite{Jaskierny:2006sr}.  Thus, prior to filling with liquid argon, the ambient atmosphere in the cryostat is removed by purging the tank with argon gas.  

\begin{figure}[htp]
\begin{center}
\includegraphics[width=0.45\textwidth]{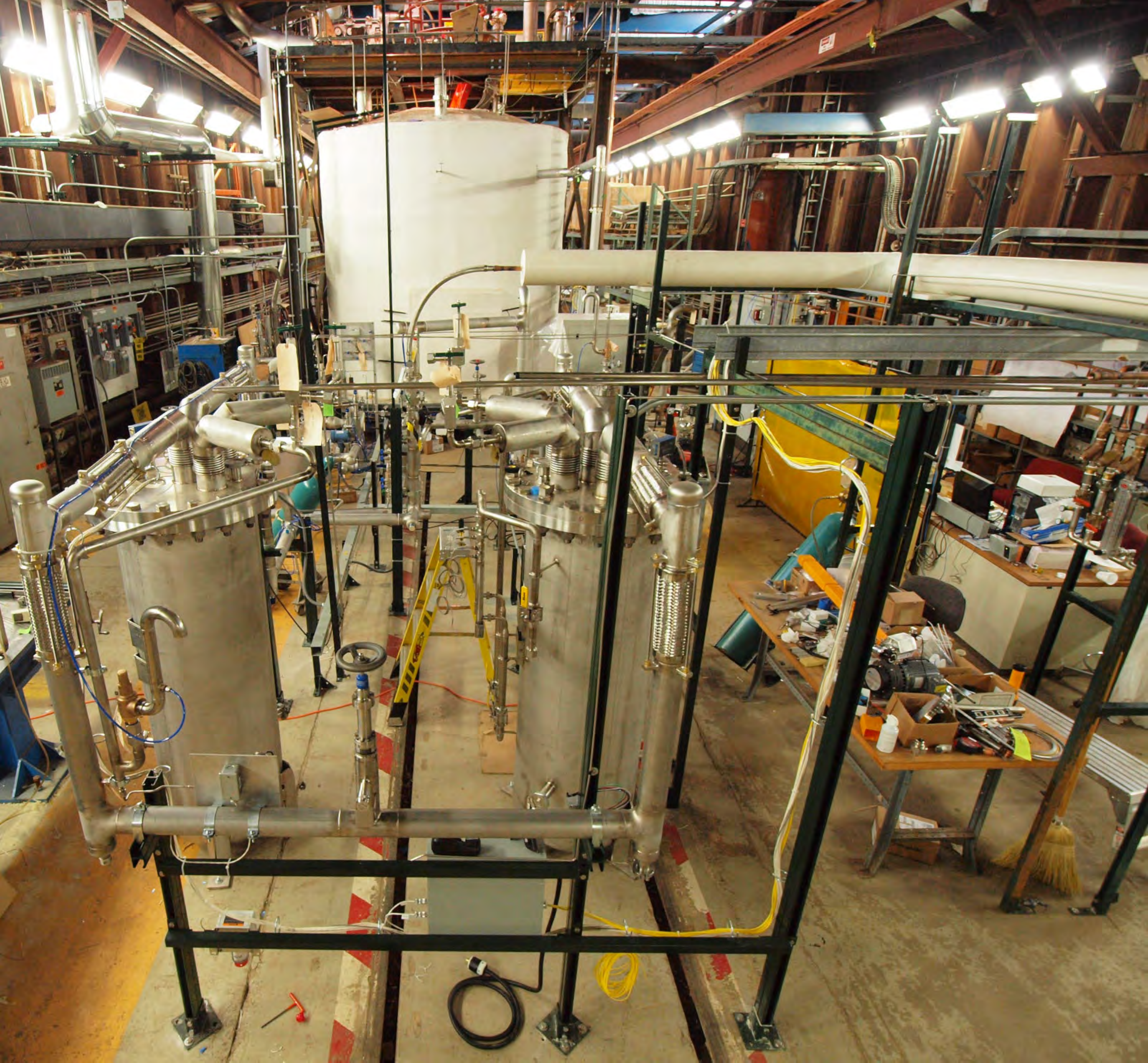}
\includegraphics[width=0.45\textwidth]{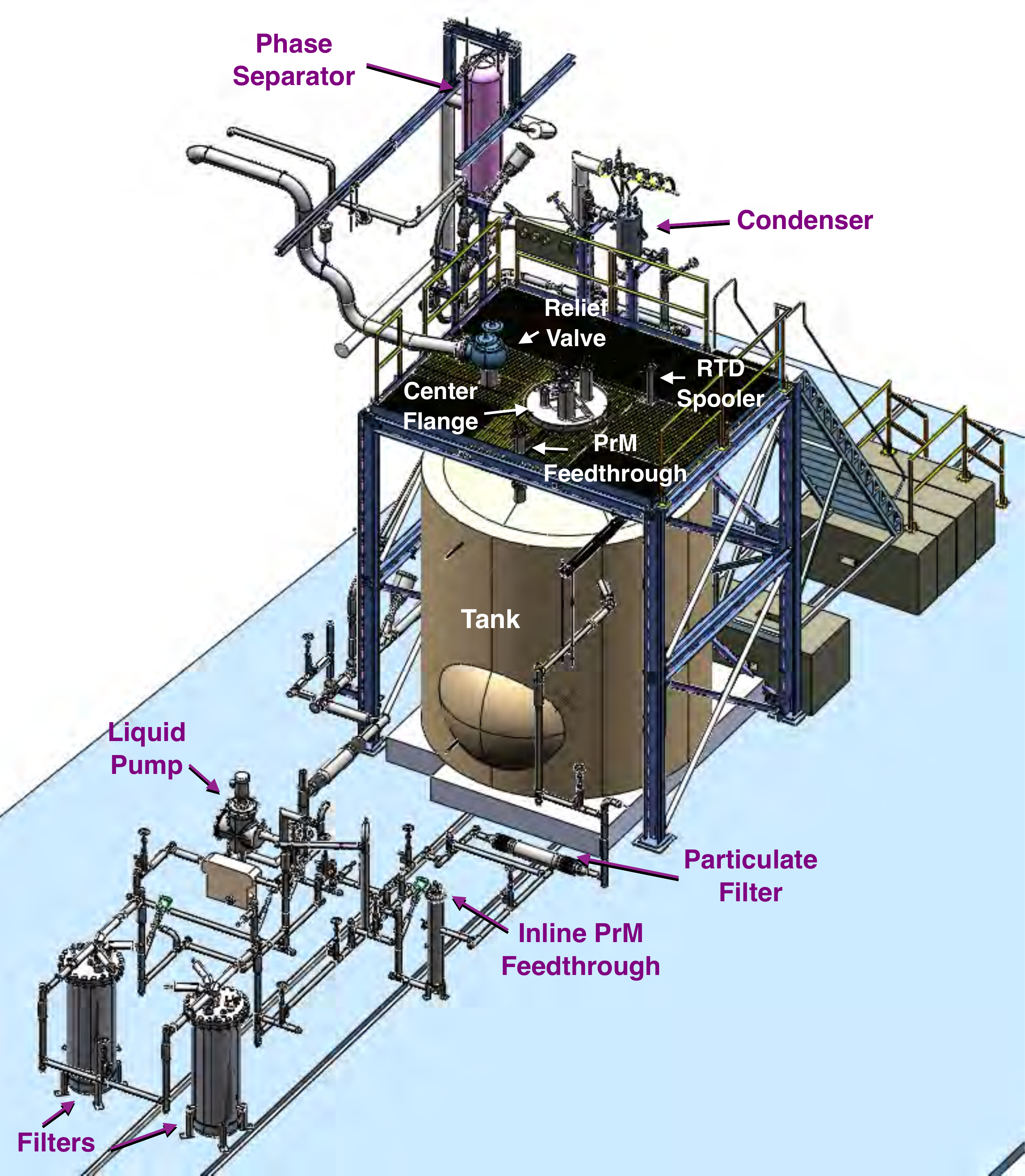}
\caption{A photograph of the Liquid Argon Purity Demonstration (LAPD) at Fermilab (left) and 3D model of the system (right). }
\label{figure:lapd}
\end{center}
\end{figure}

After the initial purge, once the water and oxygen concentrations are at the level of a few ppm, the argon gas is subsequently circulated through filters to further reduce these contaminants.  Liquid argon is then introduced into the filters after impurity concentrations less than 1 ppm are achieved. The liquid is then continuously circulated through the filters to achieve concentrations of water and oxygen on the order of 0.1 ppb. One may choose to heat the walls of the cryostat in an attempt to dry those surfaces while circulating the argon gas through the filters.  However, our tests indicate that such a step does not influence the resulting contamination levels when liquid is introduced into the vessel.  A photograph and a 3D rendering of the LAPD vessel and piping configuration is shown in Figure~\ref{figure:lapd}.

The LAPD had several secondary goals.  First, we studied the number of liquid argon volume exchanges necessary to achieve drift distances on the scale of 2.5 meters.  Second, we compared simulations of liquid argon temperature gradients and impurity concentrations in the cryostat to our measurements using dedicated instruments installed in the cryostat.  Third, we monitored and evaluated filter capacity performance as a function of flow rate.  Finally, after achieving the required electron drift lifetimes, the LAPD cryostat was emptied and a TPC of 2 m drift distance was installed in the central cryostat region.  High liquid argon purity was achieved with the TPC in the tank and the details of these results will be presented in a forthcoming paper.

\section{The Cryostat}\label{sec:tank}
\begin{figure}
\centering 
\includegraphics[width=0.5\textwidth]{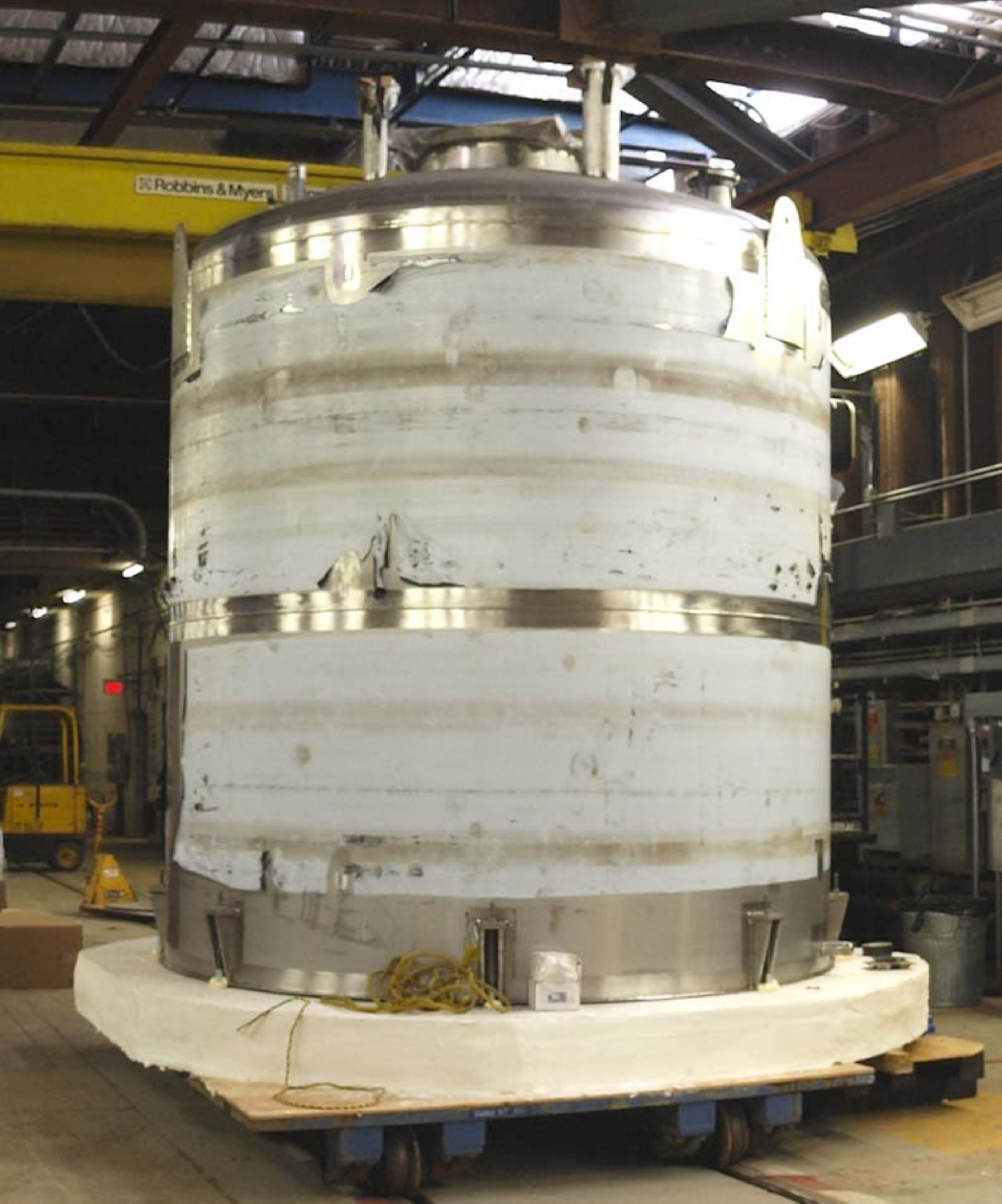}
\caption{LAPD cryostat sitting on an insulating foam base in PC4. Insulating foam was added to the sides and head later.}
\label{figure:lapdtank}
\end{figure}

The LAPD cryostat is an industrial low pressure storage tank.  The cryostat has a flat bottom, cylindrical sides, and a dished head. The cryostat diameter is 3.0 m and the cylindrical walls have a 3.0 m height.  The cryostat is fabricated from 4.76~mm-thick SA-240 stainless steel.  The internal and external (vacuum) maximum allowable working pressures are 3 psig and 0.2 psig, respectively.  Eight perimeter anchors tie the cryostat bottom to the ground to prevent cryostat uplift.  The cryostat volume is 24,628 liters of which 21,210 liters is liquid (29.7 tons) with a corresponding liquid depth of 2.9 m.  Fabrication followed The American Petroleum Institute Standard 620 Appendix Q as closely as possible and the cryostat welds were fully radiographed.  The cryostat is located inside the Proton Center 4 (PC4) building at Fermilab. Figure~\ref{figure:lapdtank} shows a photograph of the LAPD cryostat.
    
\begin{figure}
\centering 
\includegraphics[width=0.8\textwidth]{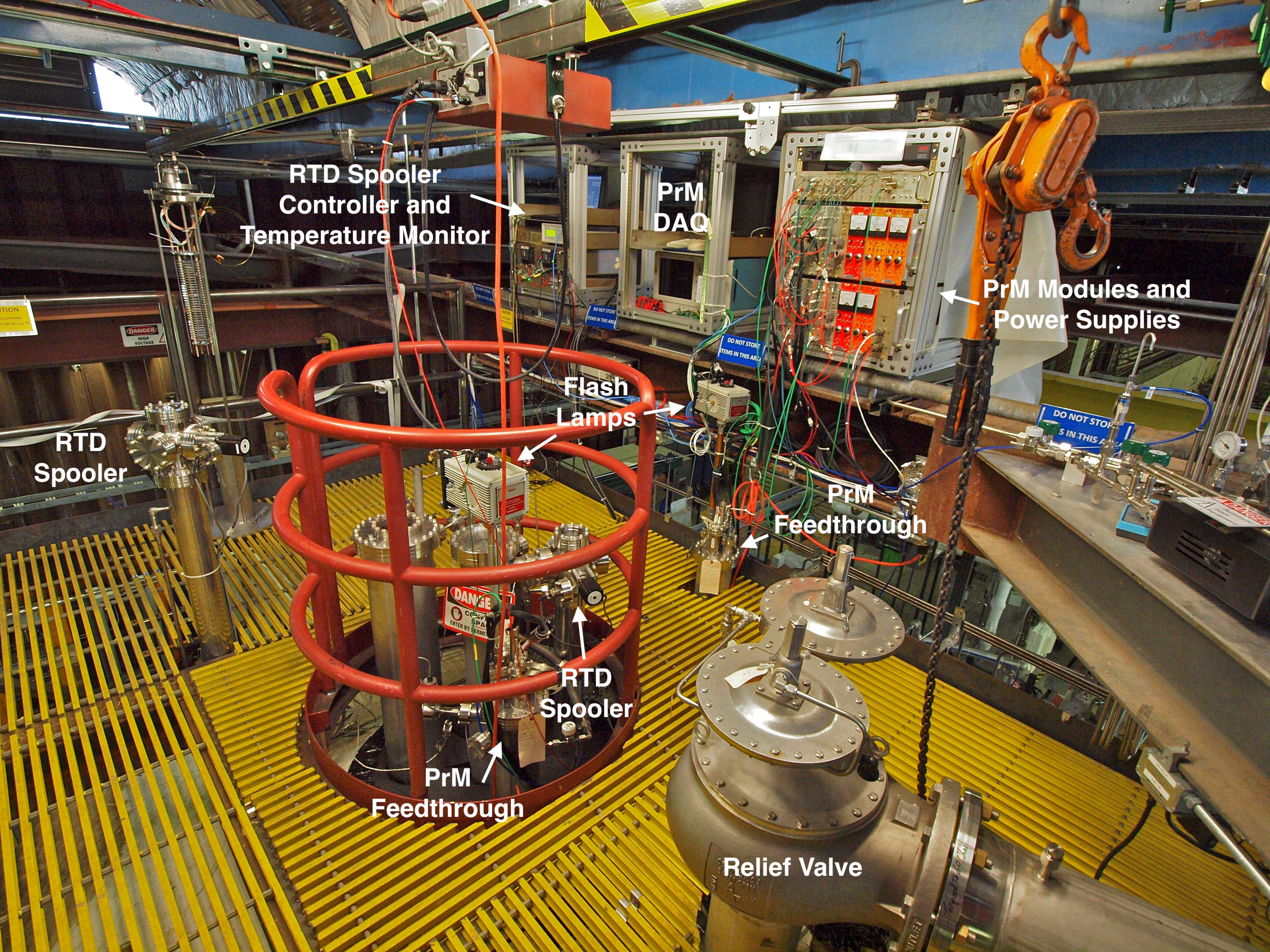}
\caption{A photograph of the platform on top of the cryostat showing the control systems (see Section~3.5), the RTD translators (see Section~4.1), and the purity monitor feedthroughs (see Section~4.2).}

\label{tanktop}
\end{figure}

The head of the cryostat is populated with four ConFlat flanges and a 76 cm diameter center flange sealed with an indium wire.  Metallic seals are used to prevent the diffusion of contamination that would occur through non-metallic seals.  The center flange allows for cryostat entry using an extension ladder.  Five ConFlat flanges populate the center flange, each of which sit atop stainless steel tube risers such that the flanges remain at room temperature when the cryostat is cold. Figure~\ref{tanktop} shows the layout at the top of the cryostat. At ground level a 76 cm diameter welded manhole is available and intended to make access easier for extended work inside the cryostat.  Table~\ref{table:op_parameters} lists the cryostat operating parameters including the heat leak, volume, operating pressure, and nominal pump flow rates.

\begin {table*}
\centering
\caption{The nominal operating parameters for the LAPD including the heat leak, operating pressures, cryostat volume, and condenser cooling capacity. }
\begin {tabular} {lr}
\hline
\hline
Designed cryostat heat leak  & 2100~W   \\
Internal max. pressure           & 3 psig   \\
External max. pressure         & 0.2 psig \\
Cryostat volume                    & 24628 liters  \\
Liquid argon volume              & 21210 liters  \\
Depth at full capacity             & 2.9 m \\
Condenser cooling capacity  & 8400 W    \\
\hline
\end {tabular}
\label{table:op_parameters}
\end{table*}

The cryostat sides and top are insulated with 25 cm of fiberglass which is covered by an outer layer of 2 cm-thick-foam.  The foam is covered with glass cloth and a layer of mastic which provides a vapor barrier.  The cryostat sits on an insulating structural foam base also sealed with a mastic vapor barrier.  The cryostat heat leak was estimated to be approximately 2100~W.  Natural air flow under the structural base eliminates the need for foundation heaters.    

The cryostat was cleaned with deionized water and detergent then dried with lint free rags by the cryostat fabricator prior to shipment to Fermilab.  After installation of all components at Fermilab, the cryostat was vacuumed with a HEPA filter equipped vacuum.  After vacuum cleaning, all walls were wiped with deionized water and lint free rags.

\section{The Cryogenics}\label{sec:cryo}
\subsection{Phase Separator and Condenser}

The argon vapor generated by ambient heat input is condensed using liquid nitrogen.  A trailer supplies liquid nitrogen through foam-insulated 2.54 cm Type K copper piping.  A phase separator operating at 15 psig near the LAPD cryostat vents nitrogen vapor generated in the nitrogen transfer line so that the condenser is supplied with single phase liquid nitrogen. The phase separator and condenser were designed at Fermilab.  A control valve feeding the phase separator maintains a constant liquid level in the phase separator.  The condenser consists of an argon volume containing three differently sized coils of tubing through which liquid nitrogen flows.  The coiled nitrogen tubing is seamless and all nitrogen connections and welds are outside the condenser to mitigate any nitrogen leak into the LAPD cryostat.  Argon vapor is condensed by the liquid nitrogen flowing through the coils. 

\begin{figure}
\centering 
\includegraphics[width=0.5\textwidth]{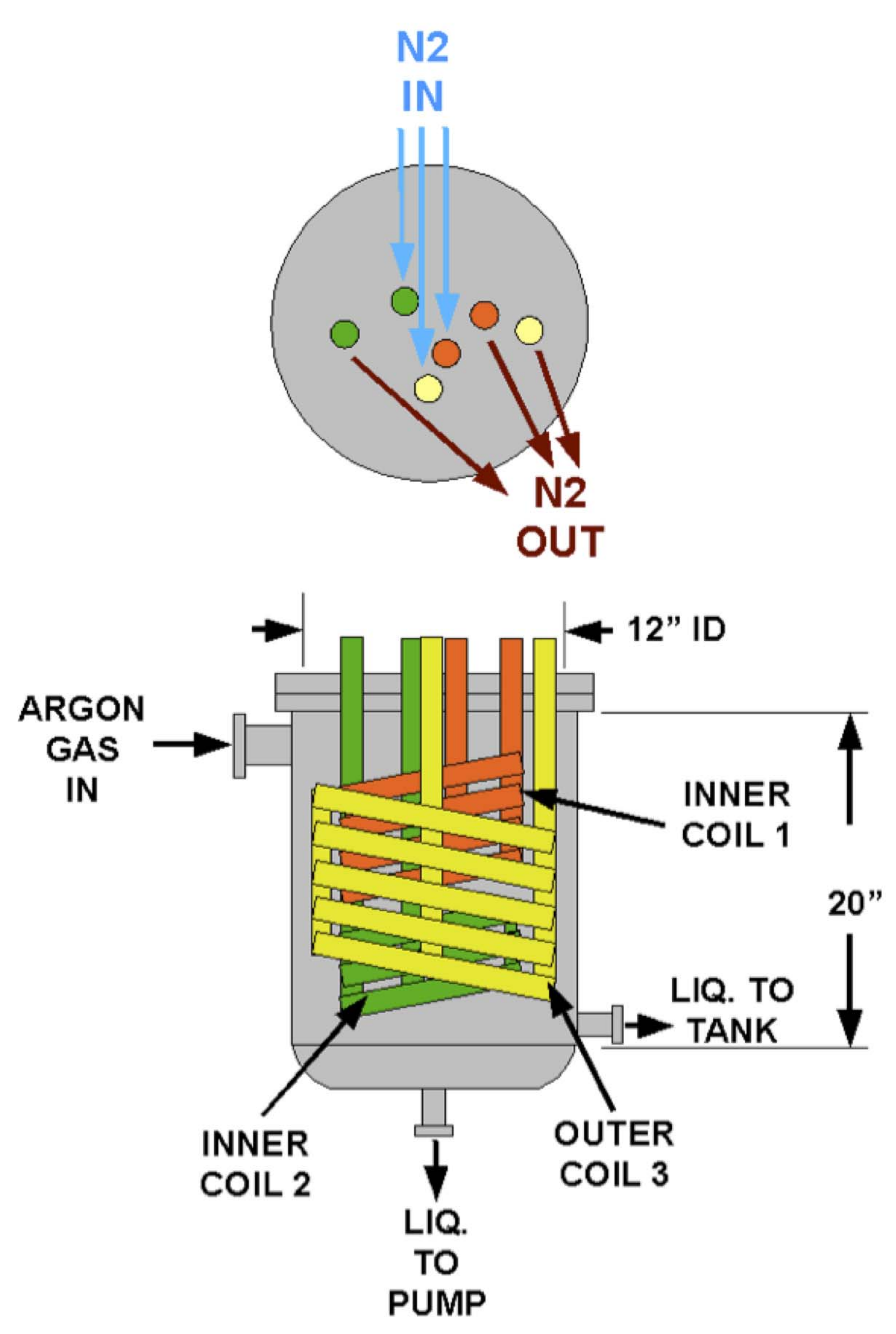}
\caption{A sketch of the LAPD condenser which shows the three coils of tubing for liquid nitrogen, the inlet for gaseous argon and two outlets for liquefied argon.}
\label{condenser}
\end{figure}

\begin{figure}
\centering 
\includegraphics[width=0.6\textwidth]{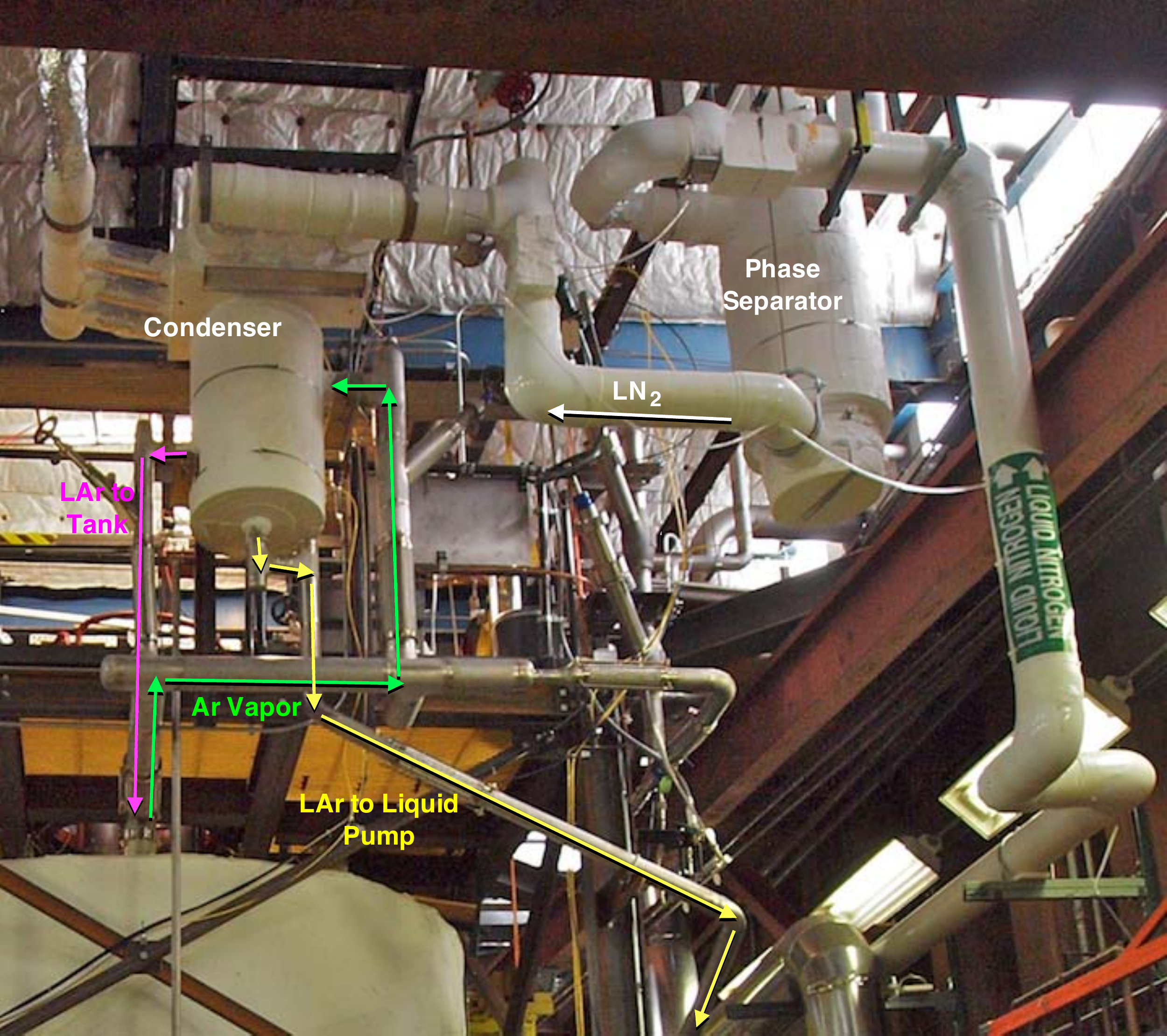}
\caption{A photograph of the LAPD phase separator and condenser. The argon vapor path to the condenser and the two liquefied argon return paths are shown.}
\label{condenserphoto}
\end{figure}

Water outgassing from the tank walls, devices, and cables above the liquid is mixed with argon vapor which needs to be removed to maintain high liquid argon purity.  Thus, by default the condensed liquid argon returns to the liquid recirculation pump suction before going through the filters during liquid recirculation. When the pump is off, the condensed liquid argon returns directly to the tank.  A control valve feeds the condenser and adjusts the flow to maintain a constant vapor pressure in the ullage.  Solenoid valves choose which combination of coils receives liquid nitrogen.  The coils operate at near ambient pressure due to the pressure drop across the inlet control valve.  The coils will therefore be covered in a thin layer of argon ice due to the large temperature gradient.  Argon ice formation was accounted for in the condenser design and no noticeable impact on the cooling due to the argon ice was observed.  Vaporized nitrogen is vented outside the enclosure and not recovered.  Figure~\ref{condenser} shows a sketch of the condenser design and Figure~\ref{condenserphoto} shows a photo of the phase separator and condenser in PC4.

\subsection{Filters}\label{sec:filters}

\begin{figure}
\centering 
\includegraphics[width=0.53\textwidth]{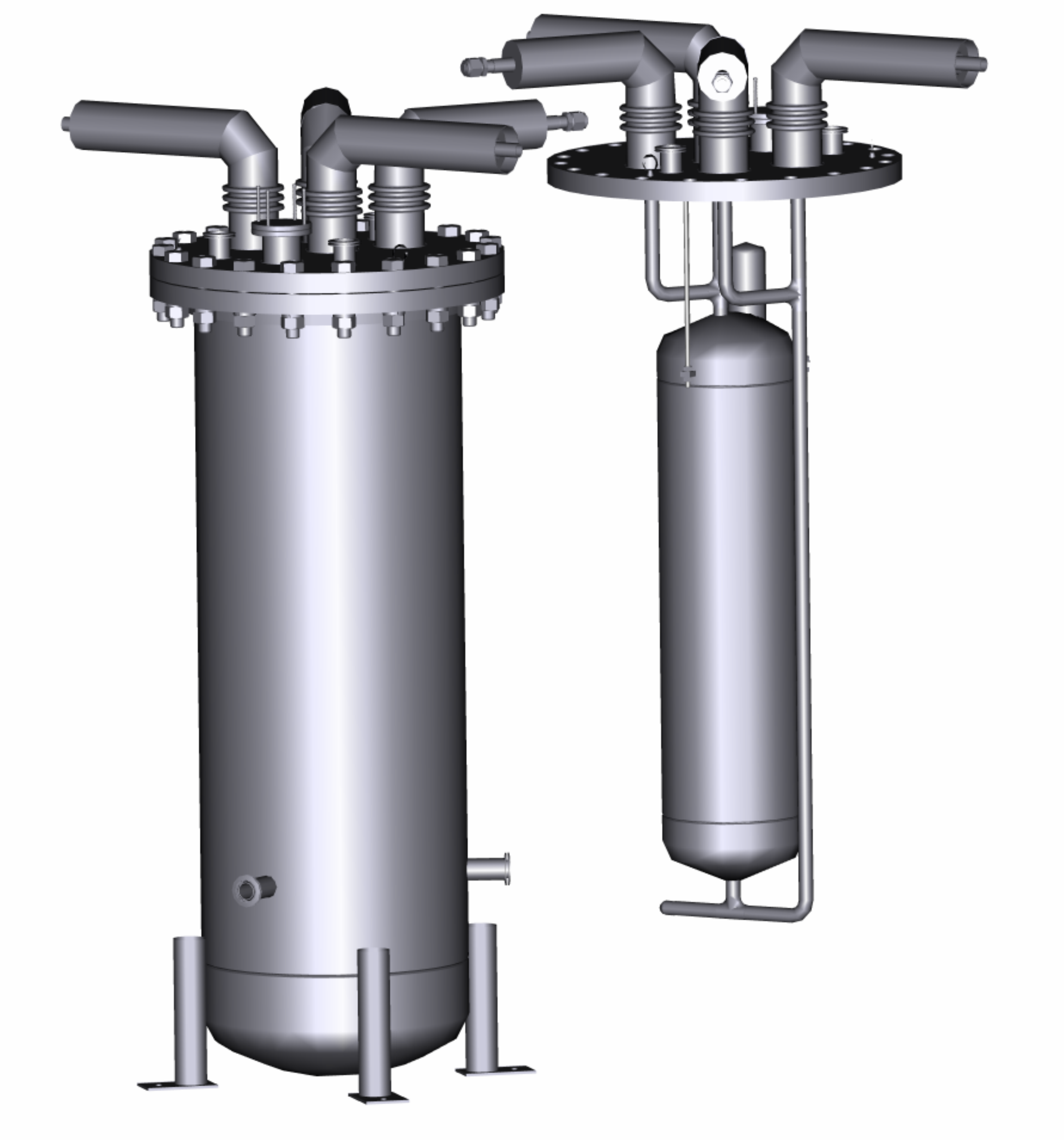}
\includegraphics[width=0.37\textwidth]{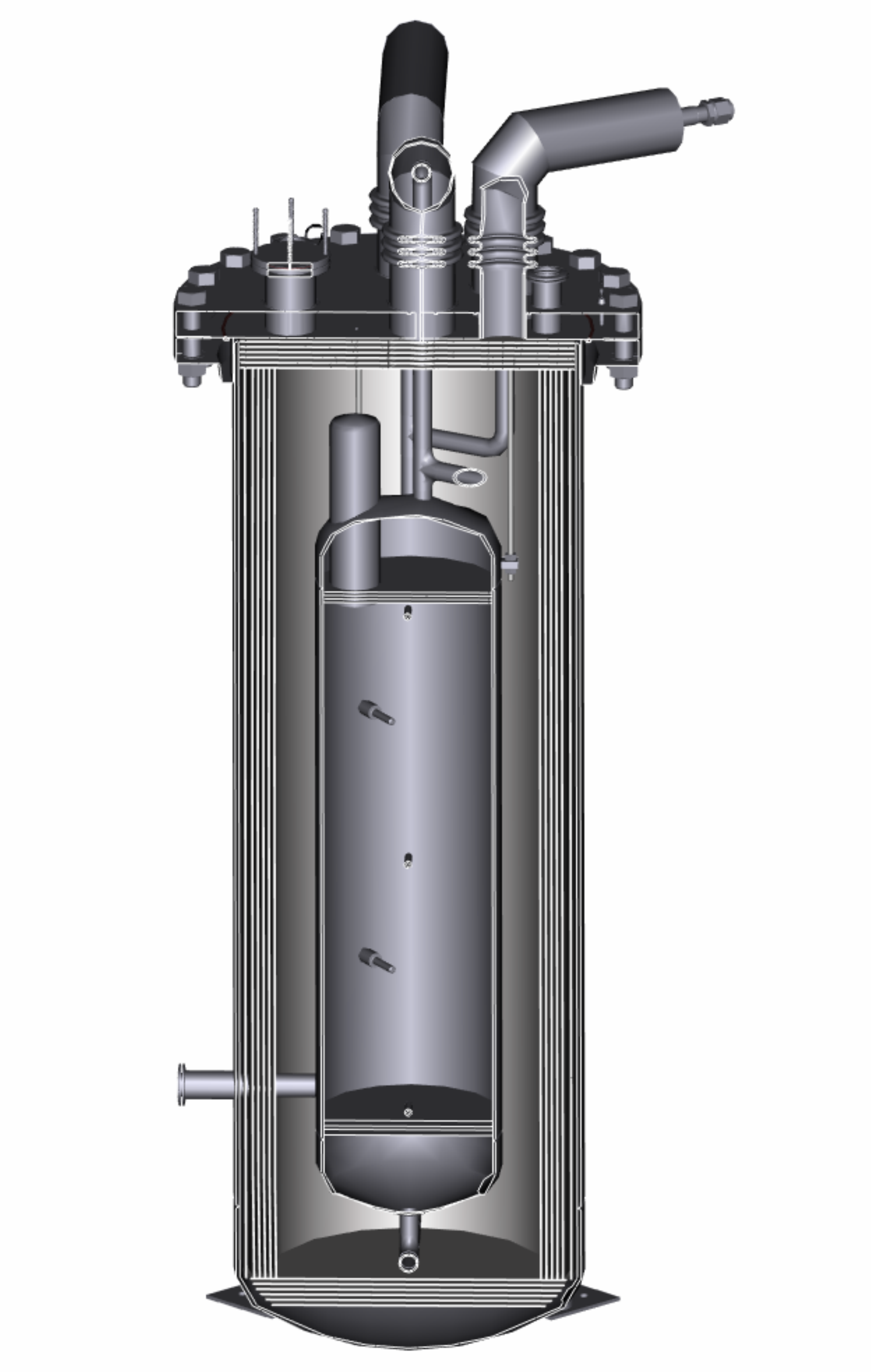}
\caption{A 3D rendering of the filter vessel. The evacuation vessel and the canister holding the filter material are shown on the left. The cross section of the vessel is shown on the right.}
\label{filter}
\end{figure}

The purification system contains two filters having identically sized filtration beds of 77 liters.  The first filter that the argon stream enters contains a 4A molecular sieve supplied by Sigma-Aldrich~\cite{sigma-aldrich} that primarily removes water contamination but can also remove small amounts of nitrogen and oxygen.    The second filter contains BASF~CU-0226~S, a highly dispersed copper oxide impregnated on a high surface area alumina, to remove oxygen~\cite{basf} and to a lesser extent, water.  Thus, the oxygen filter is placed downstream of the molecular sieve to maximize oxygen filtration.  The filters are insulated with vacuum jackets and aluminum radiation shields.  Metallic radiation shields were chosen because the filter regeneration temperatures, described below, would damage traditional aluminized mylar insulation.  Piping supplying the filter regeneration gas is insulated both inside the filter vacuum insulation space and outside the filter with Pyrogel XT which is an aerogel based insulation~\cite{aspen-aerogels} that can withstand temperatures up to 1200 F. Figure~\ref{filter} shows a 3D rendering of the filter vessel.

The filters are regenerated in place using heated gas, which differs from the procedure performed by ICARUS~\cite{Amerio:2004ze}.  Both LAPD filters are regenerated using a flow of argon gas that is heated to $200^{\circ}{\rm C}$, supplied by commercial 180 liter liquid argon dewars.   Once at $200^{\circ}{\rm C}$, a small flow of hydrogen is mixed into the primary argon flow and exothermically combines with oxygen captured by the filter to create water.  Too much hydrogen mixed in with the primary argon flow induces temperatures that are sufficiently high to damage the copper-based filter media.  The damage is induced by sintering of the copper which reduces the available filter surface area.  Thus, precautions are taken to maintain a hydrogen fraction below 2.5\% of the heated gas mixture.  During the heated gas regeneration, five filter bed temperature sensors monitor the filter material temperature and the water content of the regeneration exhaust gas is measured.  Both filters are evacuated using turbomolecular vacuum pumps while they cool to remove remaining trace amounts of water.

At the filtered liquid return to the tank, a particulate filter with an effective filtration of 10 microns protects the tank from any debris in the piping.  The filter consists of a commercial stainless steel sintered metal cylinder mounted in a custom cryogenic housing and vacuum jacket.  Filtration is accomplished by flowing liquid argon to the interior, then outward through the walls, of the sintered metal cylinder.  Flanges on the argon piping, along with flanges and edge welded bellows on the vacuum jacket, allow removal of the particulate filter.

\subsection{Piping and Valves}   

The schedule 10 stainless steel purification piping that supplies argon to the filters is vacuum jacketed.  The inner line containing argon is 2.54 cm in diameter with a 7.62 cm diameter vacuum jacket, except at the pump suction where the inner line is 5.1 cm in diameter with a 12.7 cm diameter vacuum jacket.  During the fabrication process, all piping was washed with deionized water and detergent to remove oil and grease then cleaned with ethanol.  All valves associated with the argon purification piping utilize a metal seal with respect to ambient air either through a bellows or a diaphragm to prevent the diffusion of oxygen and water contamination.  The exhaust side of all relief valves are continuously purged with argon gas to prevent diffusion of oxygen and water from ambient air across the o-ring seal.  Where possible, ConFlat flanges with copper seals are used on both cryogenic and room temperature argon piping.  Pipe flanges in the system are sealed using spiral wound graphite gaskets.  Smaller connections are made with VCR fittings with stainless steel gaskets.  


\subsection{Recirculation Pump}

The liquid argon pump is a Barber-Nichols~\cite{barber-nichols} BNCP-32B-000 magnetically driven partial emission centrifugal pump which isolates the pump and liquid argon from the electric motor.  The impeller, inducer, and driving section of the magnetic coupling each have their own bearings that are lubricated by the liquid argon at the impeller.  The motor is controlled by a variable frequency drive (VFD) which allows adjustment of the pump speed to produce the desired head and flow within the available power range of the motor.  

The liquid argon flow rate is measured at the pump discharge by an Emerson Process Management Micro Motion Coriolis flow meter~\cite{emerson}.  This flow meter is appropriate for ultra high purity liquid argon because, from the perspective of the liquid argon, it only consists of stainless steel pipe and flanges.  The inertial effects of the fluid flow through the flow meter pipes is directly proportional to the mass flow of the liquid.  The mass flow rate is computed by measuring the difference in the phase vibration between the two ends of the flow pipe.  The flow curve of the liquid argon pump with respect to mass flow and pressure is relatively flat, such that pump speed and differential pressure are not good indicators of the mass flow rate.  Thus the liquid argon flowmeter is essential instrumentation if the rate of filtration is to be known.  

\begin{figure}[Htb]
\centering 
\includegraphics[width=0.8\textwidth]{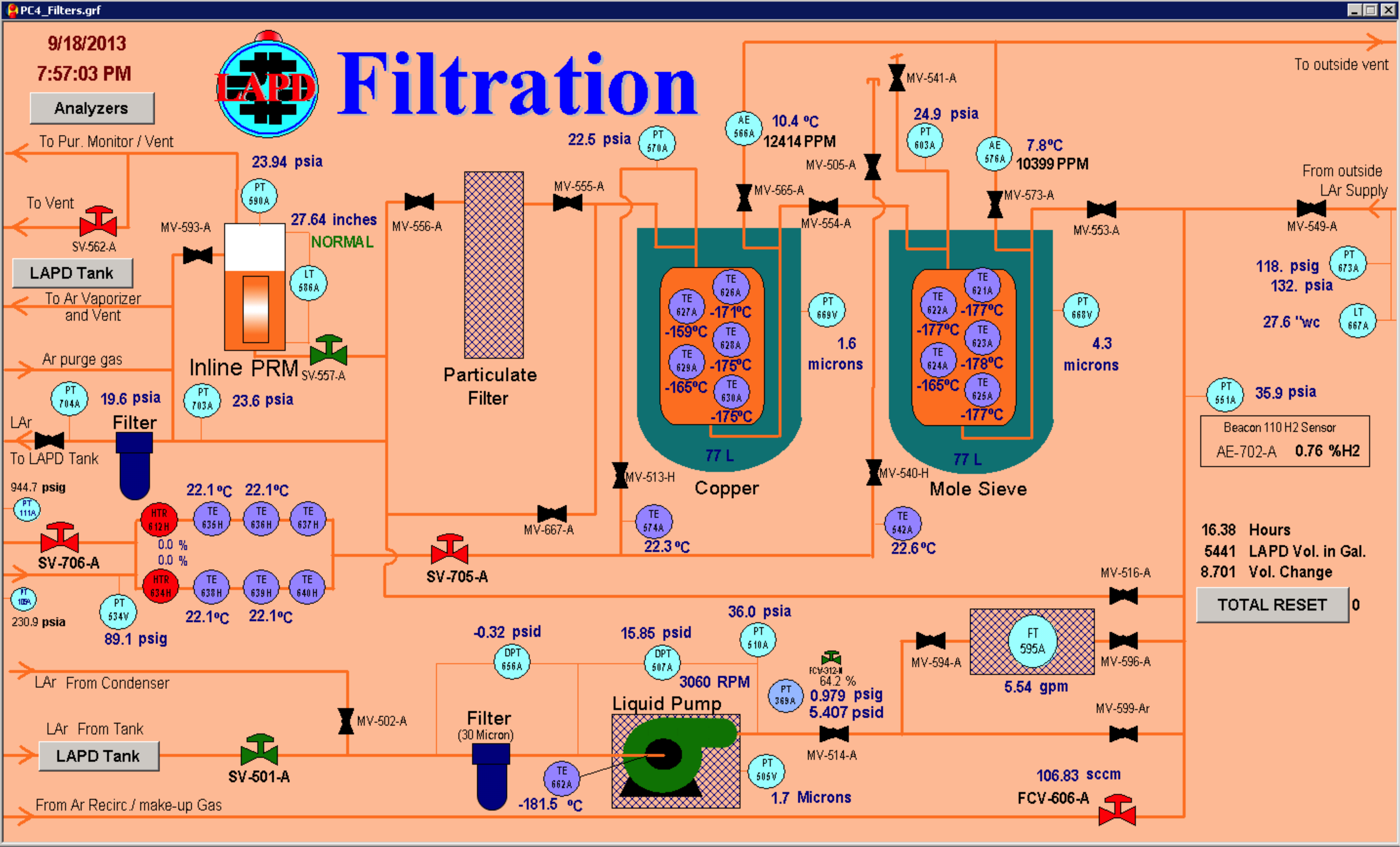}
\caption{An example of the iFIX graphical user interface for the LAPD controls.}
\label{figure:plc}
\end{figure}

\subsection{Control System}\label{ssec:control_system}

The LAPD cryogenic system is controlled by a Siemens Programmable Logic Controller (PLC)~\cite{siemens}.  The PLC reads out the pressure, liquid level, temperature, gas analyzer instrumentation, and electron lifetime measured by purity monitors.  Human-machine interface controls are provided through iFIX software~\cite{ifix} running on a PC, which is connected to the PLC through local ethernet.  The iFIX software allows entry of temperature and pressure set points and other operational parameters, handles alarming and remote operator controls such as opening and closing valves, displays real-time instrument values, and archives instrument values for historical viewing.  An example of the iFIX graphical user interface used in the LAPD is shown in Figure~\ref{figure:plc}.

\section{Cryostat Instrumentation}\label{sec:instrumentation}

\begin{figure}
\centering 
\includegraphics[width=0.7\textwidth]{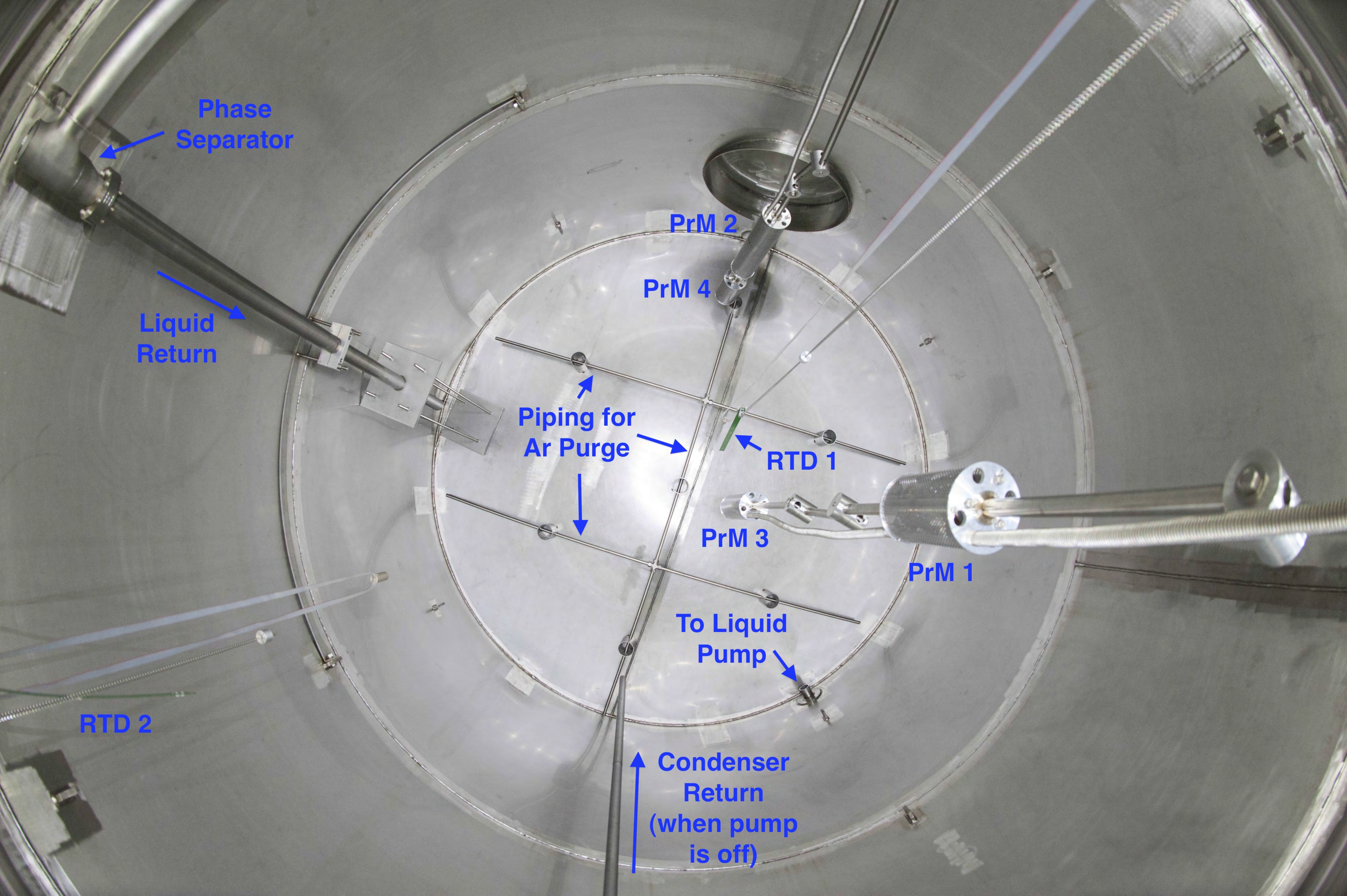}
\caption{The interior of the cryostat viewed from the top. The purity monitors, RTD translators and piping are shown.}
\label{prmrtdinLAPD}
\end{figure}

\subsection{RTD Translators}\label{ssec:translators}

Two sets of three resistive thermal devices (RTDs) on translators were deployed to measure thermal gradients in the cryostat at all stages of operation and argon fill level.  The motivation for installing these translators is to verify finite element analysis (FEA) calculations used to model liquid argon mass flow in the cryostat and to monitor potential convective flow or boiling. 

\begin{figure}
\centering 
\includegraphics[width=0.7\textwidth]{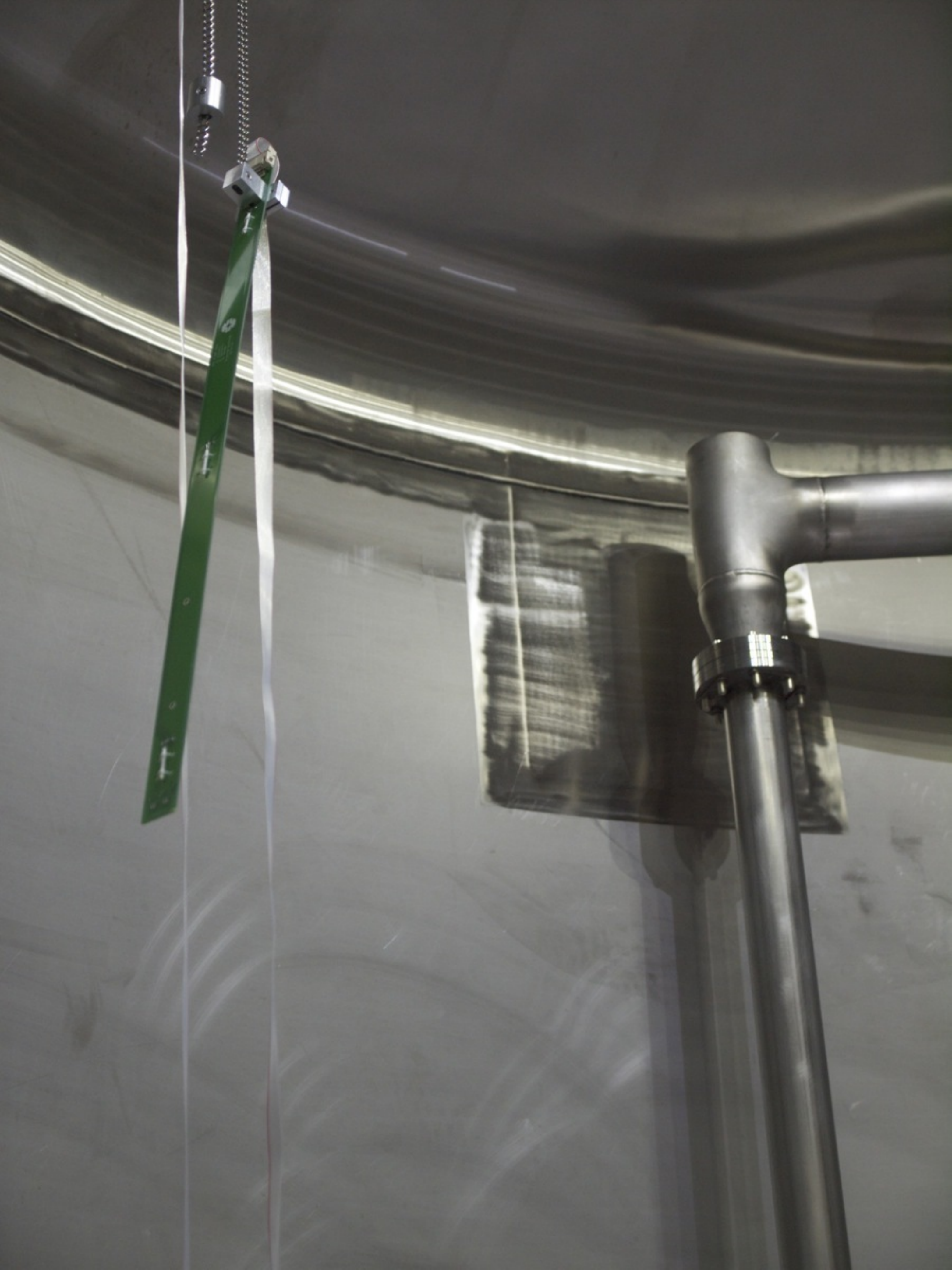}
\caption{The peripheral RTD translator inside the LAPD cryostat.}
\label{spoolerinLAPD}
\end{figure}

One translator is installed near the center of the cryostat and the other is installed 1.0~m radially outward from the center.  Figure~\ref{prmrtdinLAPD} shows the locations of the RTD translators inside the cryostat as well as the purity monitors, discussed in Section 4.2. Each translator consists of a 50 cm-long circuit board with three RTDs mounted at 22.9 cm intervals, as shown in Figure~\ref{spoolerinLAPD}.  The circuit board is suspended at one end of a chain, with a counter-weight at the other end of the chain to prevent movement during an electrical outage.  The chains engage a 15.13 cm circumference  gear that is driven externally, through a ferromagnetic seal, by an Automation Direct STP-MTRH-23079 stepper motor~\cite{automationdirect}. The housing around the gear also includes electrical limit switches to stop the motor when the chain limits are reached. The stepper motor is controlled by an Automation Direct STP-DRV-4850 stepper drive~\cite{automationdirect}. 

Teflon ribbon cables connect the circuit boards to a LakeShore model 218 temperature monitor which reads out all six RTDs.  The stepper motor controller and LakeShore are controlled and read out by a custom LabVIEW application.  The RTDs are platinum, type K 100 Ohm and were measured to be accurate to within 0.5 K.  During a typical acquisition, the circuit board translates vertically through the cryostat and temperature data are acquired at twenty equidistant steps spanning the cryostat height.  Each data point consists of sixty-four single RTD measurements with times between steps being long enough to mitigate bias in movement and ensure thermal equilibrium.  Acquisition times between data points are approximately 5 and 30 minutes for measurements in the liquid and gas, respectively.

\subsection{Purity Monitors}\label{ssec:purityMon}

The purity of liquid argon is continuously monitored by a double gridded ion chamber immersed in the liquid argon volume.  The fraction of electrons generated at the cathode that arrive at the anode ($Q_{A}/Q_{C}$) after the electron drift time, $t$, is a measure of the electronegative impurity concentration and can also be interpreted as the electron lifetime, $\tau$ such that

\begin{equation}
Q_{A}/Q_{C} = e^{-t/ \tau}.
\end{equation}

\subsubsection{Hardware}

The purity monitor is based on the design described in Reference~{\cite{Carugno:1990kd}. It consists of four parallel, circular electrodes: a disk supporting a photocathode, two open wire grids, one anode and one cathode, and an anode disk. The anode disk and photocathode support disk are made of stainless steel; the two grid support rings are made of G-10 circuit board material with the grid wires soldered to the copper clad surface. The region between the anode grid and cathode grid contains a series of field-shaping stainless steel rings. The two grids, each with a single set of parallel wires, are made of electro-formed gold-sheathed tungsten with a 2.0 mm wire spacing, 25 $\mu$m wire diameter and 98.8\% geometrical transparency.

The cathode grid is at ground potential. The cathode, anode grid, and anode are electrically accessible via modified vacuum grade high voltage feed-throughs. The anode grid and the field-shaping rings are connected to the cathode grid by an internal chain of 50 M$\Omega$ resistors to ensure the uniformity of the electric fields in the drift regions. To ensure maximum transparency~\cite{Bunneman}, the field ratios typically satisfy

\begin{equation} 
E_{1} > 2 E_{2} > 4 E_{3},
\end{equation}

\noindent where $E_{1}$ is the field between the anode grid and the anode, $E_{2}$ is the field between the anode grid and the cathode grid, and $E_{3}$ is the field between the cathode and the cathode grid.

The photocathode is a $2.54$~cm~$\times~3.81$~cm~$\times~0.8$~mm aluminum plate, coated with 50 \AA\ of titanium and 1000 \AA\ of gold and attached to the cathode disk~\cite{platypus}. A xenon flash lamp~\cite{newport} is used as the light source. The UV output of the lamp has a wide spectrum above approximately 225~nm. An inductive pickup coil on the power leads of the lamp provides a trigger signal when the lamp flashes. Light is directed via three multi-mode quartz optical fibers~\cite{polymicro} to the photocathode. Only one fiber is needed; the other two are for redundancy. The fibers have a 0.6~mm core diameter and 25.4 degrees of full acceptance cone. The attenuation is 0.95 dB/m at 200 nm. The fibers underwent a series of tests using a photodiode read out by an oscilloscope to measure the stability and light output linearity as a function of input light intensity and showed no anomalous behavior.

The electrons liberated from the photocathode drift towards the cathode grid and induce a current on the cathode. After crossing the cathode grid, the electrons drift between the two grids. During this time essentially no current is induced on the cathode or anode due to the shielding effect of the grids. After crossing the anode grid, the electrons induce a current on the anode. The signals induced on the cathode and anode are fed into two charge amplifiers in a purity monitor electronics module. The charge amplifiers have a 5 pF integration capacitor with a 22 M$\Omega$ resistor in parallel with the capacitor. The signal and high voltage are carried on the same cable and decoupled inside the purity monitor electronics module. Figure~\ref{figure:prmschematic} shows a schematic of a purity monitor installed in the cryostat.

\begin{figure}[htp]
\begin{center}
\includegraphics[width=0.45\linewidth]{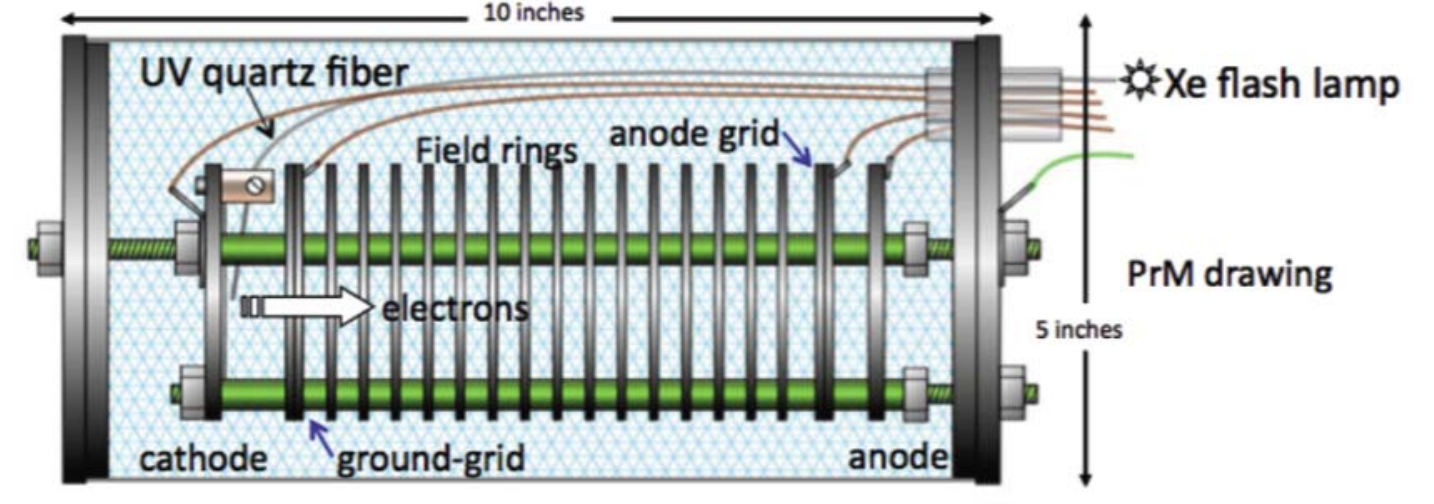}
\includegraphics[width=0.45\textwidth]{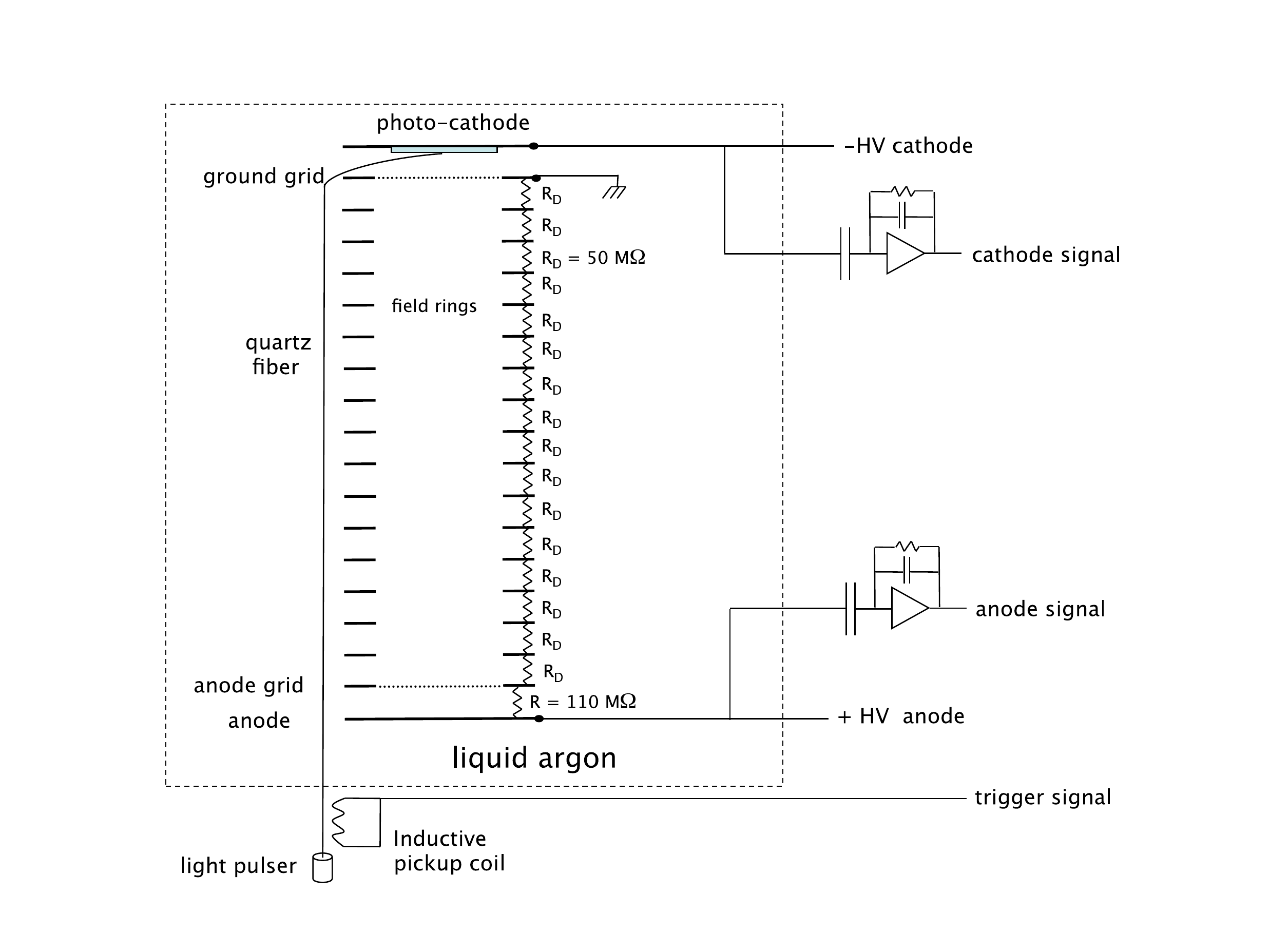}
\caption{A drawing (left) and schematic (right) of a liquid argon purity monitor employed in LAPD.  }
\label{figure:prmschematic}
\end{center}
\end{figure}

The LAPD system employs five purity monitor units at different locations. Each purity monitor is contained in a stainless steel, perforated Faraday cage to isolate the system from outside electrostatic interference. There are two types of purity monitors with different lengths and different numbers of field-shaping rings: three long purity monitors that are 55 cm in length and two short purity monitors that are 24 cm in length.  The operational range of $Q_{A}/Q_{C}$ for which a purity monitor can make sensible measurements is about 0.05 - 0.95.  Thus, longer electron drift lengths correspond to operational ranges shifted to sample larger electron lifetimes.  An assembly of one long purity monitor and one short purity monitor is located vertically along the central axis of the cryostat. Another identical assembly is located at a distance of 1.1 m away from the center of the cryostat. Figure~\ref{prminLAPD} shows a photograph of the assembly located near the cryostat periphery.  One long purity monitor, referred to as the inline purity monitor, is located in the circulation pipe to measure the liquid argon purity before the liquid enters the cryostat. Three flash lamps are used for the two purity monitor assemblies and the inline purity monitor. Table~\ref{table:prmgeo} shows the geometrical characteristics and voltage settings of the purity monitors installed in the cryostat.

\begin{figure}
\centering 
\includegraphics[width=0.45\textwidth]{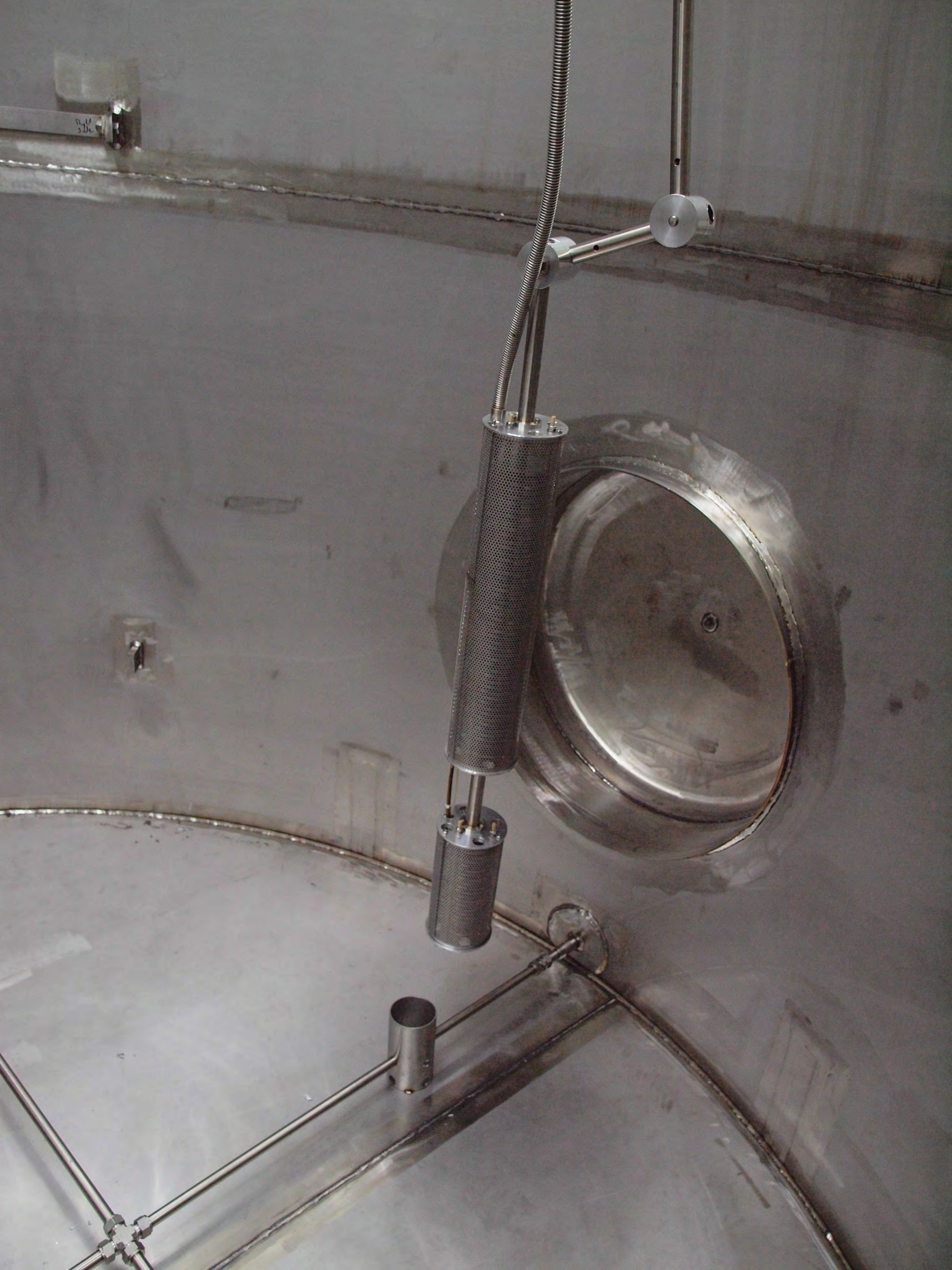}
\caption{An assembly of one long purity monitor and one short purity monitor located near the cryostat wall inside LAPD.}
\label{prminLAPD}
\end{figure}

\begin{table}[htp]
\begin{center}
\caption{Geometrical characteristics and voltage settings of the purity monitor. $V_{AG}$ and $V_{A}$ are the applied voltages on the anode grid and at the anode, respectively. Electric field values are reported for nominal operation.}
\begin{tabular}{l c c } \hline\hline
&Long monitor & Short monitor\\ \hline
Cathode, Anode disk, grid diameter & \multicolumn{2}{c}{8 cm}\\
Cathode-Cathode Grid gap & \multicolumn{2}{c}{1.8 cm}\\
Anode Grid-Anode gap & \multicolumn{2}{c}{0.79 cm}\\
Anode disk/Cathode disk thickness & \multicolumn{2}{c}{0.23 cm}\\
Anode grid/Cathode grid thickness & \multicolumn{2}{c}{0.24 cm}\\
Field-shaping ring thickness & \multicolumn{2}{c}{0.23 cm}\\
Gap between rings & \multicolumn{2}{c}{0.79 cm}\\
Cathode-Anode total drift distance & 50 cm & 19 cm\\
Cathode grid to Anode grid distance & 47 cm & 16 cm\\
Number of field-shaping rings & 45 & 15\\
Number of resistors & 46 & 16\\
Nominal Cathode Voltage & -100 V & -100 V\\
Nominal Anode Voltage & 5 kV & 2 kV\\
V$_{AG}$/V$_{A}$ &0.948&0.865\\ 
Electric field between cathode grid and cathode & 56 V/cm & 56 V/cm\\
Electric field between cathode and anode grids & 101 V/cm & 108 V/cm\\
Electric field between anode and anode grid & 329 V/cm & 342 V/cm\\
\hline\hline
\end{tabular}
\end{center}
\label{table:prmgeo}
\end{table}

\subsubsection{Data Acquisition}

\begin{figure}
\begin{center}
\includegraphics[width=0.6\textwidth]{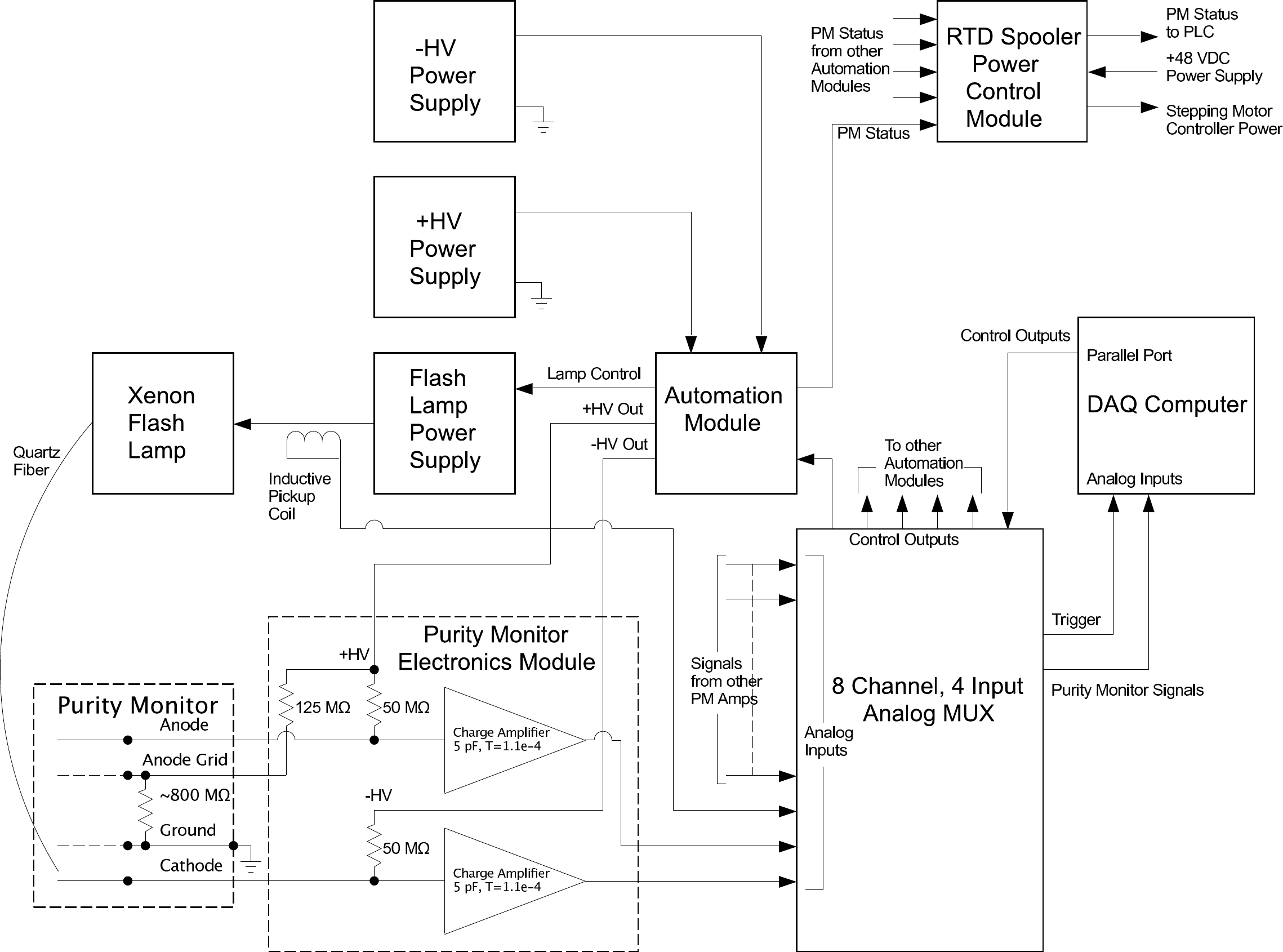}
\caption{Block diagram of LAPD purity monitor system.}
\label{figure:prmblock}
\end{center}
\end{figure}

Measurements of the electron lifetime are taken several times a day using a Fermilab-designed data acquisition (DAQ) program. Each measurement takes about one minute. The flash lamp and the high voltage to the purity monitors are only powered during this time to protect the flash lamp, minimize degradation of the quartz fiber and reduce dust/particle accumulation on the purity monitor photocathode. The automation module will switch off both the flash lamp power supply and high voltage to the purity monitor if the lamp has been flashing for more than 140 seconds.  An 8-channel analog multiplexing unit (MUX) is used to select which purity monitor signal is readout. Each channel of the MUX has four inputs, three of which read the cathode and anode signals from one purity monitor after the amplifiers and the trigger signal from the inductive pickup coil.  Figure~\ref{figure:prmblock} shows a block diagram of LAPD purity monitor system.

The program initializes and reads out the signals from each purity monitor one by one.  When the flashlamp fires, a large voltage perturbation is induced on the cathode and anode connections which distorts the shape of the electron signals.  This perturbation is measured and recorded before turning on the high voltage to the purity monitors.  The anode and cathode signals from each purity monitor are then measured by constructing the average of ten waveform samples per acquisition, each of which are stored for offline analysis.  The pulses registered from the voltage perturbation due to the light source are subtracted from the signal before calculating the lifetimes.  A plot of the averaged and smoothed signal traces produced from a purity monitor, before and after noise subtraction, is shown in Figure~\ref{figure:Run2721_04}.

\begin{figure}
\begin{center}
\includegraphics[width=1.0\textwidth]{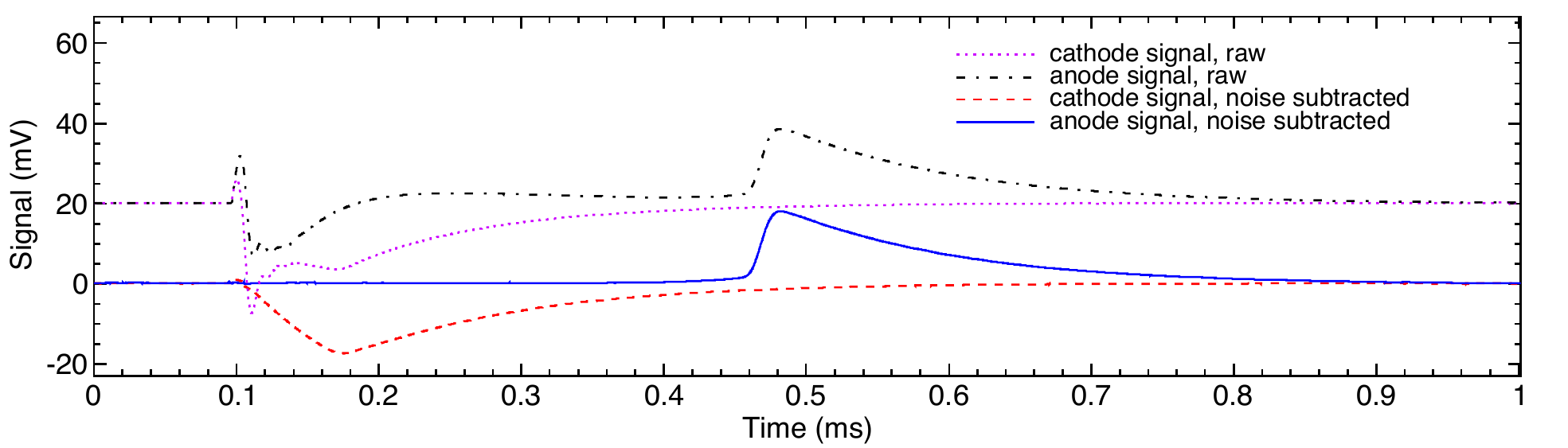}
\caption{A screenshot of anode and cathode signals before and after noise removal from the digitizer. }
\label{figure:Run2721_04}
\end{center}
\end{figure}

The maximum pulse height, $V_{max}$, of the anode and cathode traces are identified for each waveform as the maximum number in the array of numbers that stores the noise-subtracted digitized waveform.  The charge seen by the cathode and anode is then calculated as 

\begin{equation}
Q~=~(V_{max}~-~V_{0})~\times~f(\Delta~t,~RC), 
\end{equation}

\noindent where $V_{0}$ is the measured baseline for the trace and $f(\Delta t, RC)$ is a correction for the electronics response function that depends on the time duration of the current pulse, $\Delta t$, and the measured RC time constant of the electronics.  

An additional source of electrical noise that affected the operation of the purity monitor DAQ was found to be the RTD translator stepper motor controllers. These controllers have a DC to DC switching converter that provides the holding current to the stepper motors used in the RTD translator system. The most effective way to mitigate this noise source was to remove the 48 volt DC bulk power to the stepping motor controllers whenever the purity monitor DAQ was running. After the purity monitors were turned off by the DAQ, the 48 volt DC power was restored to the stepping motor controllers and a reset signal was given to the controllers so that they would index back to the zero starting point for their data collection.

\subsubsection{Electron Lifetime and Attenuation Corrections}

The electron lifetime relies on measurements of the anode voltage, $V_{A}$ and the cathode voltage, $V_{C}$, which in turn depend on amplification of induced currents on the anode and cathode. The potential exists for differences in amplification between the anode and cathode signal voltages to have an impact on the lifetime. We model the amplification as $V_{A} = g_\alpha Q_{A}$ and $V_{C} = g_\beta Q_{C}$, where $g_\alpha$ and $g_\beta$ are constants with units of ADC bins per fC. If the two amplifiers used for the anode and cathode signals are switched, the amplification becomes $V_{A}' = g_\beta Q_{A}$ and $V_{C}' = g_\alpha Q_{C}$. The primes indicate measurements taken with the amplifiers for the anode and cathode swapped.  The lifetime and attenuation calculations can then be calibrated by making measurements of $V_{A}$, $V_{A}'$, $V_{C}$, and $V_{C}'$ using

\begin{equation}
\frac{g_{\alpha}}{g_{\beta}} = \sqrt{ \frac{  V_{A}/V_{C} }   { V_{A}'/V_{C}'  } }.
\end{equation}

During a span of several days at nearly constant argon purity, measurements were taken with the amplifiers swapped to measure the ratio $ g_{\alpha}/g_{\beta}$.  For the purity monitor with the best performance, $ g_{\alpha}/g_{\beta}$ was measured to be 0.97.  With the measurements taken, a correction to the electron drift lifetimes was applied using 
\begin{equation}
\tau = \frac{t} { \ln(  (Q_{C}/Q_{A})\times  (g_{\alpha}/g_{\beta} )) }.
\end{equation}

An additional cross-check was performed by examining the effect of varying the high voltage applied to the cathode and anode.  Lifetime measurements from the short peripheral monitor were taken at anode voltages of 2~kV, 3~kV, 4~kV, and 5~kV.  Each study resulted in purity measurements consistent with those at nominal high voltage.

\subsubsection{Systematic Uncertainties}

There are three sources of systematic uncertainty that could be significant to the purity monitor measurements.  For each of these quantities, we determined a lower limit for $Q_{A}/Q_{C}$, so that a lower limit on the electron drift lifetime can be quoted.  The first is the acceptance of the anode, which we assume to be 100\%.  It is different from 100\% if any electrons generated at the cathode traverse the entire drift distance without encountering an impurity, but are not counted at the anode because they travelled too far transversely from the axis of the purity monitor.  The second is the possibility that electrons generated at the cathode induce a signal on the cathode grid, and are thus counted in $Q_{C}$, but were actually absorbed by an impurity before arriving at the cathode grid.  We assume that this absorption is negligible, and thus the value of $Q_{C}$ used is an upper limit.  The third is the uncertainty on the $RC$ time constant used to correct for the electronics response, as discussed previously.  If any of these quantities is different than assumed, then the actual values of $Q_{A}/Q_{C}$ and the electron drift lifetime are larger than quoted.

\subsection{Gas Analyzers}
\subsubsection{Oxygen, Water and Nitrogen  Monitors}
The LAPD has an extensive gas analysis system to monitor and diagnose the processes that take the cryostat from atmospheric air to ultra pure liquid argon.  The system consists of seven commercial gas analyzers.  Four of these analyzers measure the oxygen concentration and together they span the range from 0.1 ppb to 5000 ppm.  These four oxygen analyzers are augmented by two 0.1-25\% oxygen sensors which monitor the purge of the cryostat of air and are described in Section~\ref{oxygensniffers}.  Two of these seven gas analyzers measure the water concentration and these span the range from 0.2 ppb to 20 ppm.  Dew point meters installed in series with these water analyzers extend the measurement range from 20 ppm up to ambient dew points as high as 20,000 ppm water.  A nitrogen analyzer completes the array of seven gas analyzers with a range that spans 0.1 to 100 ppm.  

The gas analyzers are fed by a local switchyard of 56 diaphragm valves.  These valves direct the gas flow from five primary locations in the system to the seven gas analyzers.  A primary location or utility gas can feed anywhere between none and all of the gas analyzers.  The primary measurement locations are the liquid argon cryostat, with the option of sampling from either the gas or liquid phases, pump discharge, molecular sieve filter output, oxygen filter output, and the liquid argon fill connection.  In addition to the five primary locations, argon and nitrogen gas from utility sources are available to supply analyzers when measurement from a system location is not required.  

An oil free vacuum pump is also part of the switchyard and can evacuate the tubing that connects the measurement locations and the gas analyzers.  Evacuation of the sample lines when switching sample locations greatly reduces the time required to reach equilibrium when the measured contamination is at the parts per billion concentration.  A high purity metal bellows pump boosts the sample pressure from the 2 psig operating pressure of the liquid argon cryostat to the 15-20 psig inlet pressure required by the gas analyzers. A photograph of the gas distribution switchyard is shown in Figure~\ref{figure:switchyard}.

\begin{figure}
\begin{center}
\includegraphics[width=0.5\textwidth]{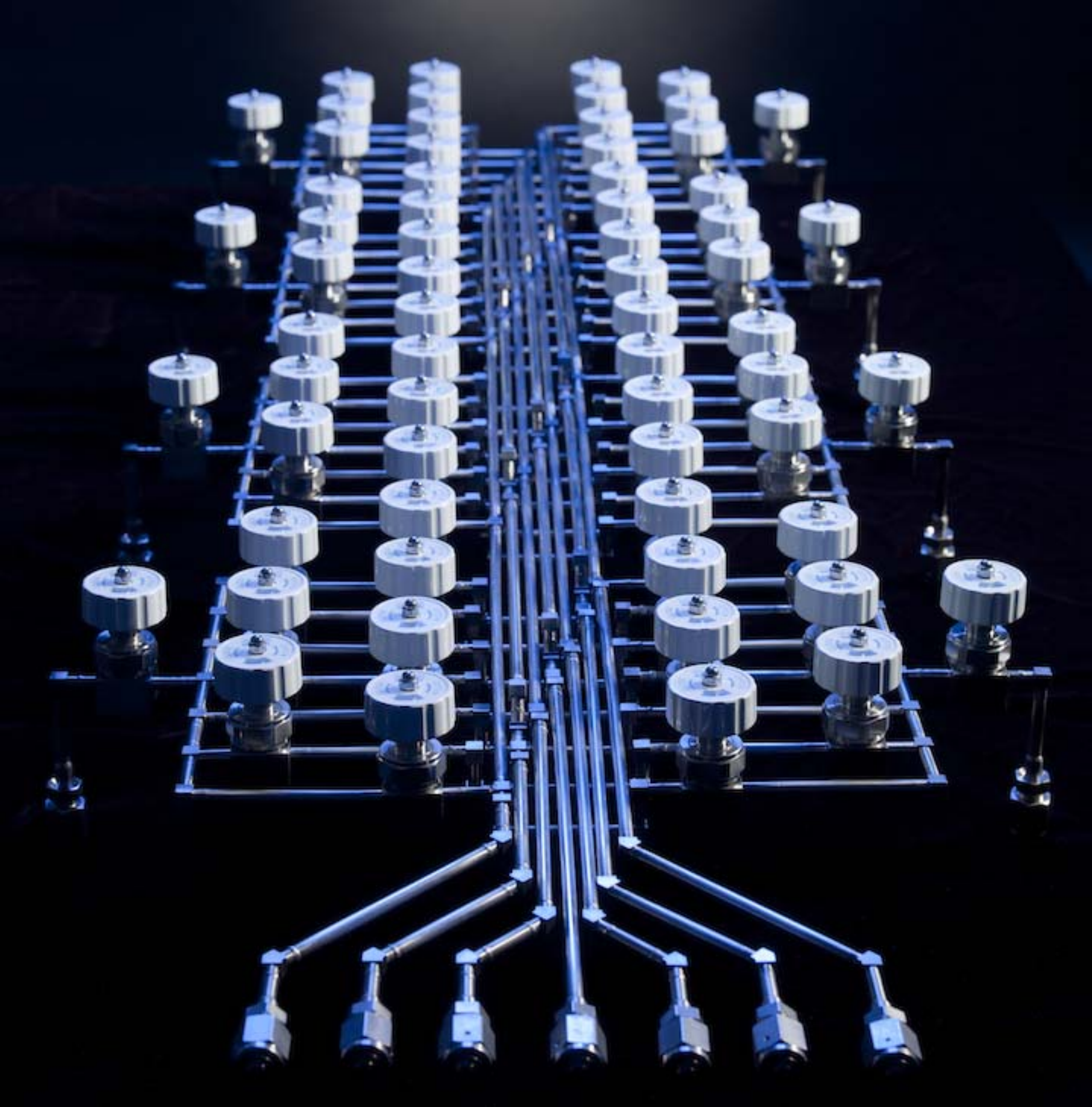}
\caption{Gas distribution switchyard - liquid argon gas sampling master distribution panel.}
\label{figure:switchyard}
\end{center}
\end{figure}

Filter output sampling allows determination of filter performance and capacity.  Sampling the liquid argon fill connection is critical to ensure that the liquid argon supply is within specification.  For example, a trailer of liquid argon was rejected because it was so far out of specification it would have required an impractical number of filter regenerations to process.  Without this extensive gas analyzer system it would be very difficult to successfully take the cryostat from ambient air to ultra pure liquid argon.

\subsubsection{Oxygen Capillary Detectors}
\label{oxygensniffers}

We deployed thirteen industrial type oxygen sensors, configured in two strings of six and seven, each consisting of capillary tubes placed a different heights, to measure the oxygen concentration during the initial gaseous argon purge.  One string was placed near the cryostat wall and the other was placed near the cryostat center.  The end of each tube is vertically spaced 76.2 cm from the next in the string, with the string spanning the height of the cryostat. The ``central'' set was placed on the tank axis, the ``peripheral'' set was 112 cm radially out.  The sensors are located inside small glass jars with plastic coated lids.  The sample tubes are 1.6 mm diameter capillaries, and run continuously from the intake point through a ConFlat flange to the jars.  All capillaries are the same length, with the excess length coiled up above the feed-through flange, to assure matched time response.  The jars are mounted on feed-through flanges.

\section{Results}\label{sec:results}

The LAPD was operated in two separate run periods.  In each period, the cryostat was operated in three phases; a gaseous argon purge, gaseous argon recirculation, and liquid recirculation. The first run period was September 2011 to April 2012.  Each of the three phases of operation were performed to test the devices and filters.  For this period, the cryostat was filled one-third full to confirm the feasibility of measuring the liquid argon purity.   The second period was from December 2012 through September 2013.  The cryostat was completely filled with liquid argon and measurements of the liquid argon purity were performed under various operating conditions.  This section describes the results for the second period, and when applicable, measurements are compared to those obtained in the first run period.

\subsection{Temperature Measurements}

The liquid argon in the LAPD cryostat is not at thermal equilibrium.  The vapor is continually being removed and condensed in an external condenser, then admixed with liquid argon drawn directly from the cryostat and then sent through purification filters before being returned to the cryostat. This evaporation of liquid argon from the surface is seen visually by a surface turbulence, and detected by the RTDs as a temperature drop between the bulk liquid below the surface and the vapor above in the ullage.

Figure~\ref{figure:rtd_ullage} shows a scan of the cryostat taken after the first fill of liquid argon for the central and peripheral RTD translators.  A very sharp temperature gradient, approximately 80~K over 50~cm is present in the ullage just above the liquid surface.  The data shown were taken only from the downward direction to avoid the thermal lag one sees when the circuit board is moving out of the liquid.

\begin{figure}
\centering 
\includegraphics[width=0.45\linewidth]{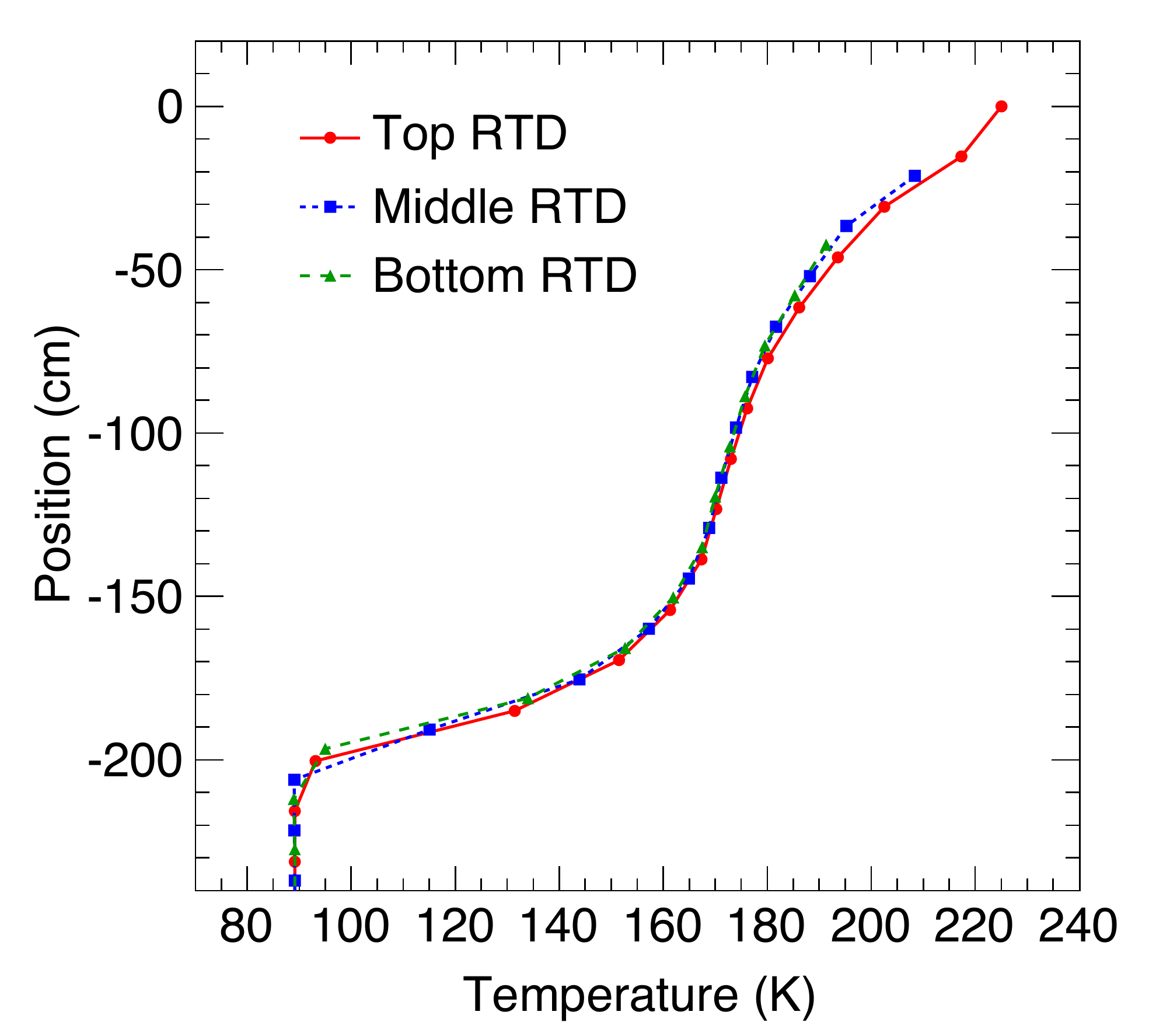}
\includegraphics[width=0.45\linewidth]{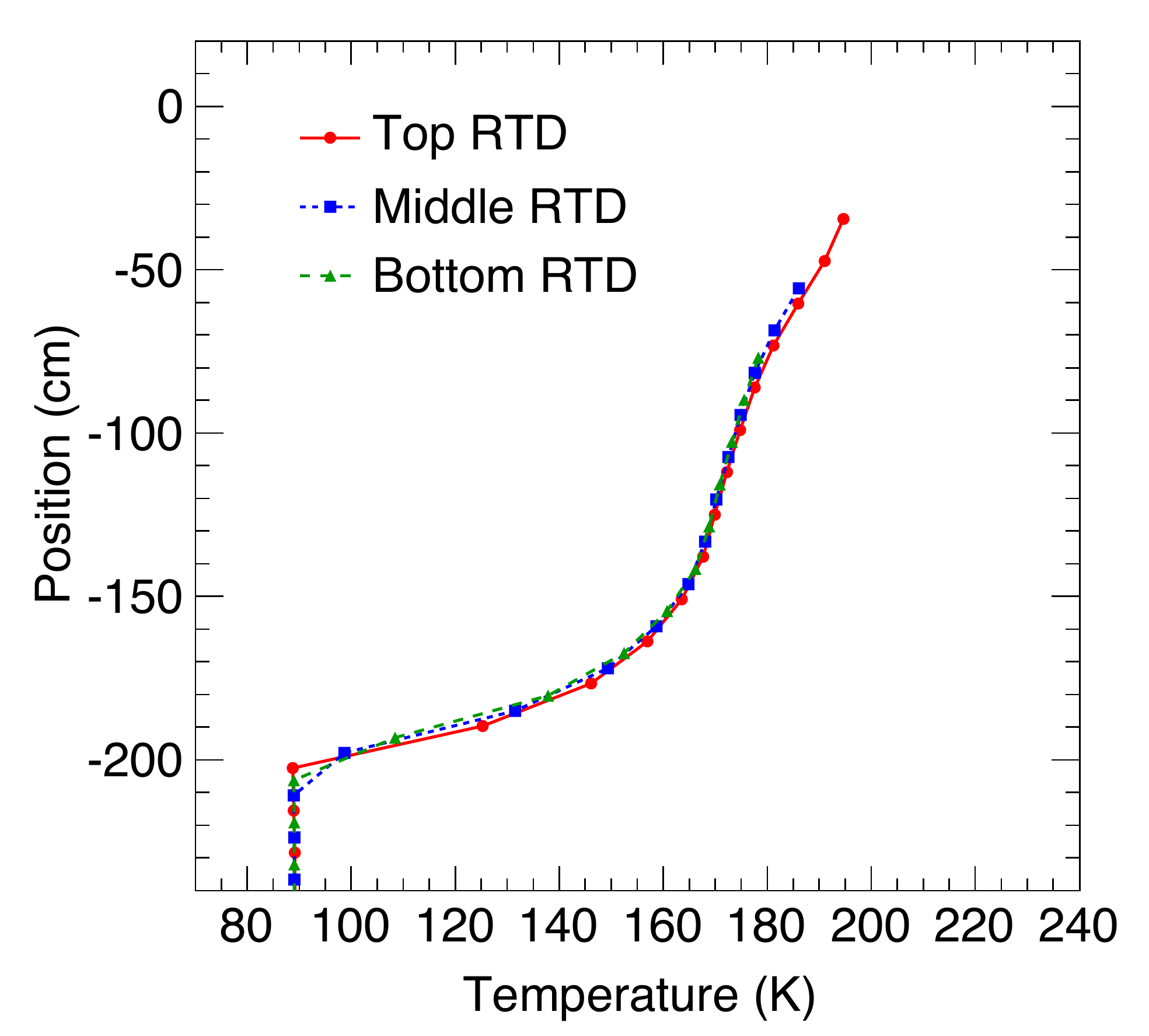}
\caption{The cryostat temperature as measured by the three central (left) and the three peripheral (right) RTDs.  Data were taken after the first trailer of liquid argon was delivered to the cryostat, corresponding to a liquid argon depth of 69 cm.  On these figures, the liquid surface corresponds to a value of -200 cm as measured from the top of the cryostat.  }
\label{figure:rtd_ullage}
\end{figure}

Figure~\ref{figure:rtd_second_results} shows the temperature measurements for the central RTDs obtained from a relatively quick scan with the cryostat full.  The slight change of temperature with depth is less than 25~mK over 2~m.  At the bottom of the scan there is a fluctuation of approximately 30~mK, attributed to the returning purified liquid argon introduced at the bottom of the cryostat.  The measurement distribution of any one RTD at a given position in the liquid, suggests a peak-to-peak spread of about ~10~mK, implying the relative precision of a single measurement is on the order of a few mK.  The measurements are compared to FEA calculations and show good agreement in both the gradient and absolute temperature.

\begin{figure}
\begin{center}
\includegraphics[width=0.45\linewidth]{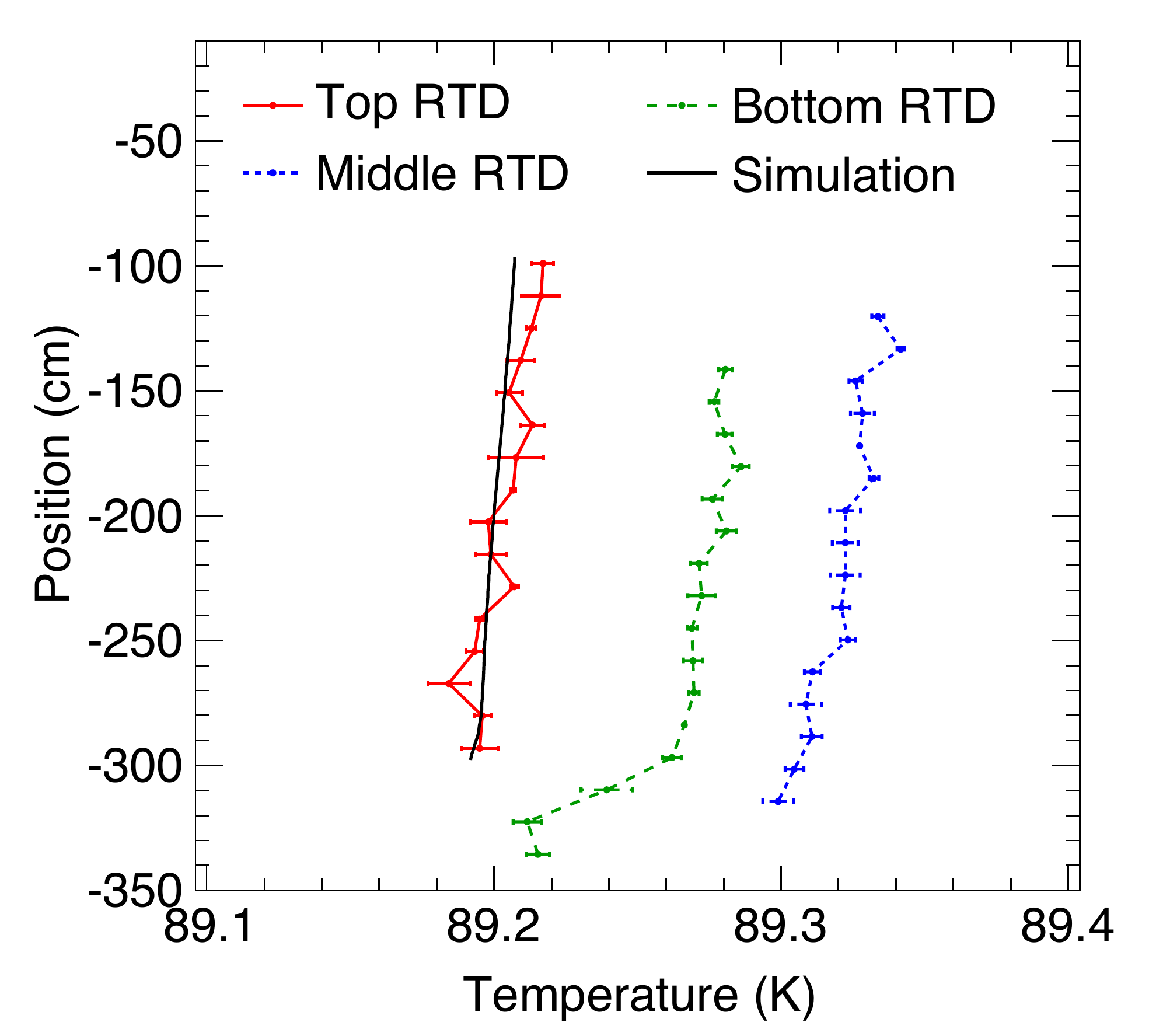}
\caption{The liquid argon temperature taken at several locations within the cryostat as measured by the three central RTDs, compared to FEA calculations.  This figure illustrates the precison with which the temperature is measured and shows a temperature variation of less than 25~mK over 2~m.  The FEA calculation is the same for all RTDs and is only compared to the top RTD position.  The precision of each RTD is 0.5~K; uncertainties reflect the distribution of measurements rather than the absolute uncertainty.   }
\label{figure:rtd_second_results}
\end{center}
\end{figure}

\subsection{Gaseous Argon Purge}

A gaseous argon purge was peformed at the beginning of each run period.  In this phase of operation, gaseous argon is pumped from the bottom of the cryostat displacing the ambient air which exits out the room temperature feedthroughs at the top of the cryostat.  This method mimics an argon ``piston'' in the sense that the higher density gaseous argon engenders a boundary between it and the ambient air, which moves vertically upwards.  For the first period, the two sets of sampling gas capillaries, described in Section 4.3.2, were installed to measure the oxygen concentration and follow the rise of the argon gas as it displaces the lighter ambient air.  

The purpose of these measurements was to understand the initial levels of impurities and obtain information for comparsion to FEA flow models to validate or improve those models.  The spatial and temporal concentration measurements provide information about the degree of diffusion and mixing during purges.  Each purge lasted approximately 8 volume exchanges or 24 hours and corresponds to a 1.15 m/hour piston rise rate and 2.9 and 3.4 hours per volume exchange for the first and second period, respectively.  The gaseous argon flow rate was constant throughout each purge.  Figure~\ref{figure:sniffer_data} shows the fraction of ambient air retained with respect to the measured oxygen levels during the gaseous argon purge in the first run period for the seven capillary tubes installed in the central region of the cryostat and the six capillary tubes installed in the peripheral region of the cryostat.  The front of gaseous argon is clearly present as indicated by the successive reduction of oxygen seen by each capillary tube as a function of time.  The FEA calculations reproduced the behavior of the argon front, but with a time lag of about 45 minutes.  This lag is suspected to be due to imperfect modeling of the capillary tube geometry.  

\begin{figure}[h]
\begin{center}
\includegraphics[width=0.45\textwidth]{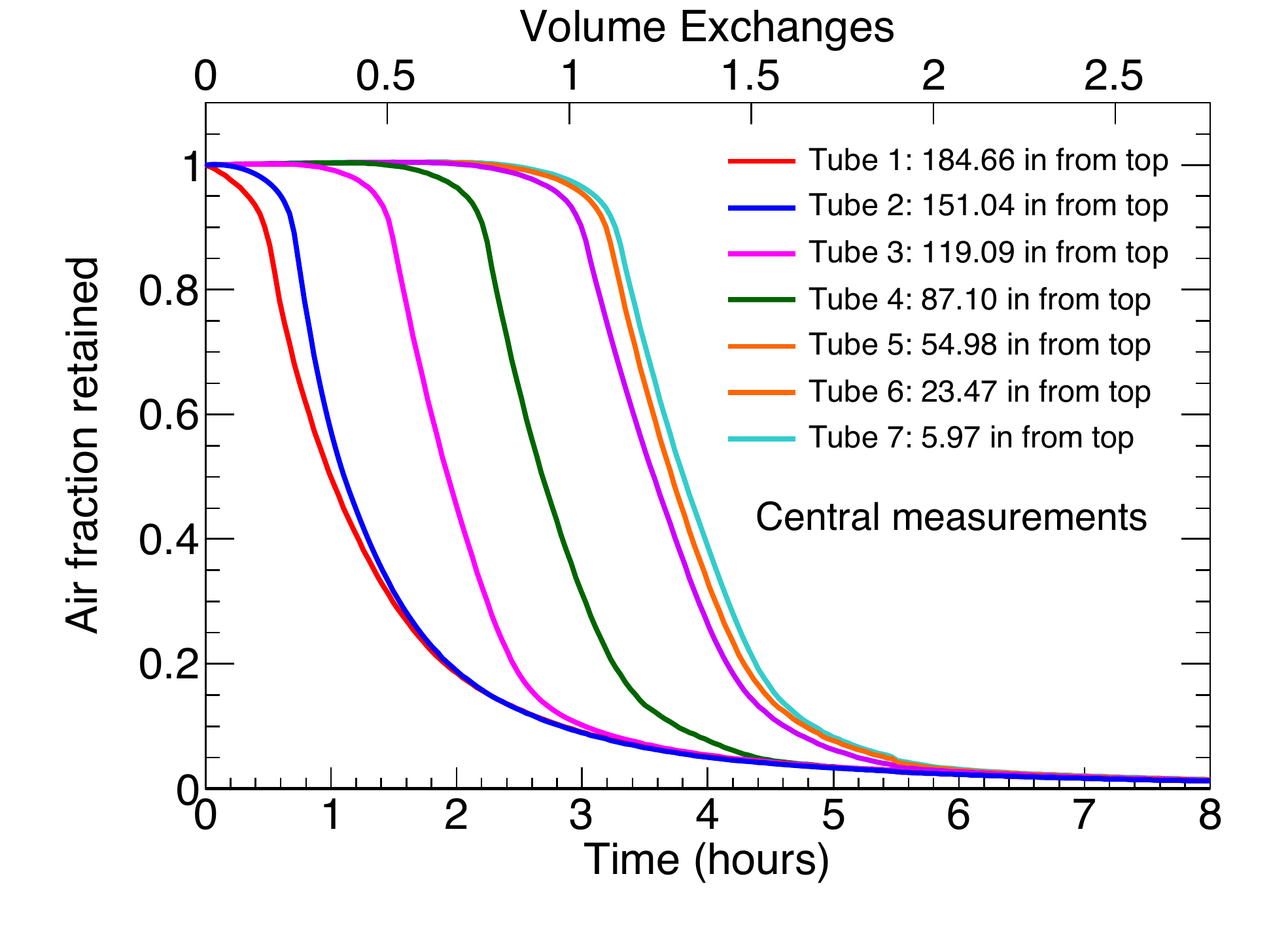} 
\includegraphics[width=0.45\textwidth]{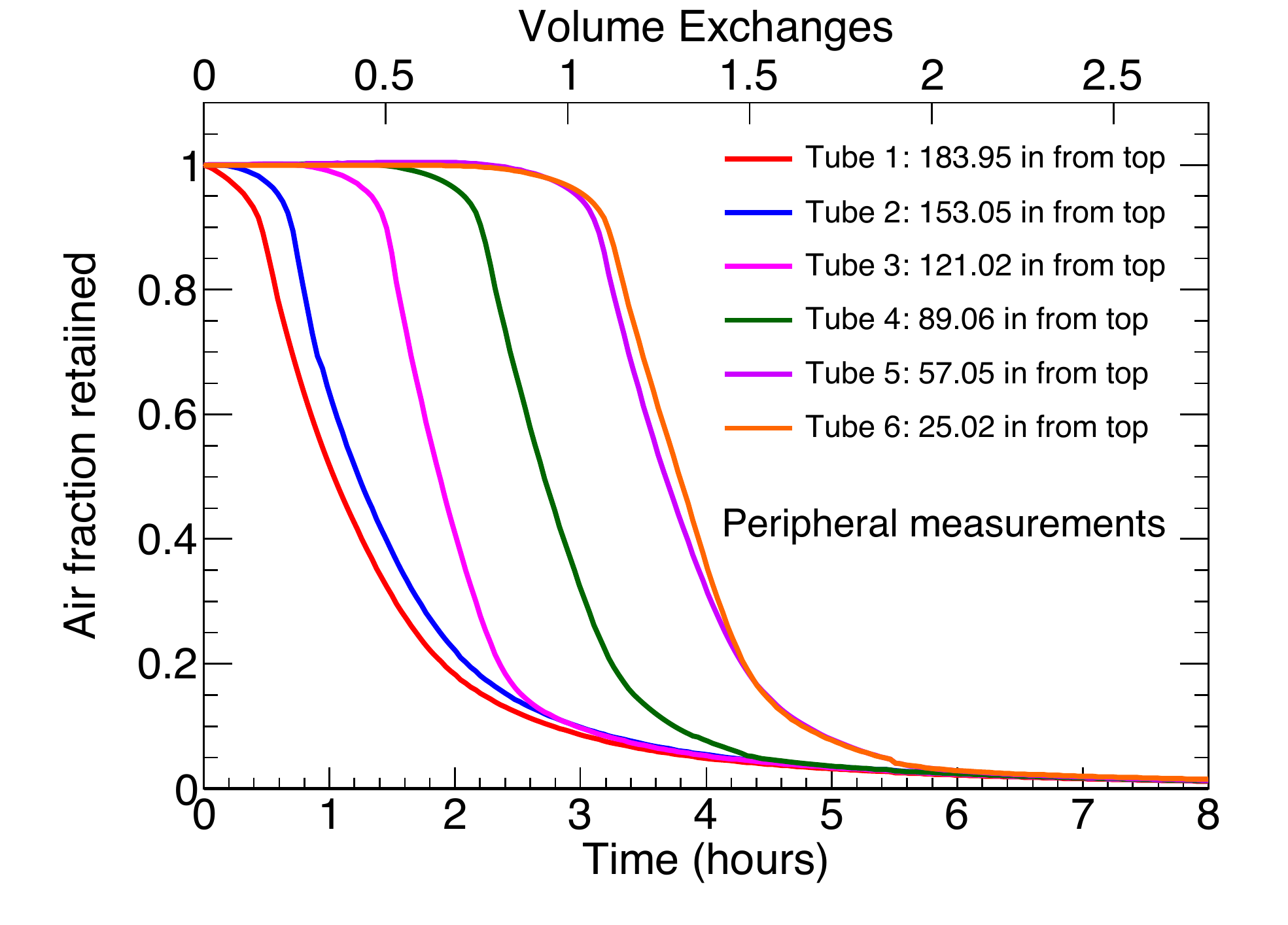}  
\end{center}
\caption{Oxygen concentrations for the a) central and b) peripheral gas sampling capillaries taken at several heights with respect to the cryostat bottom obtained during the initial gaseous argon purge for the first run period. }
\label{figure:sniffer_data}
\end{figure}

The capillaries were removed at the end of the first run period purge.  The removal lasted 15 minutes, during which time argon gas flowed into the cryostat at 0.14-0.17 cubic meters per minute (m$^3$/min).  During the extraction, the water, oxygen, and nitrogen monitors were switched to argon gas utility as a precaution because the bellows pump drawing gas from the cryostat could pull a vacuum on the cryostat if the argon flow into the cryostat stopped.  After these devices were switched back to measuring the cryostat gas, an increase of about 0.2 ppm in oxygen concentration and 0.4 ppm nitrogen concentration was observed.  With the capillaries removed, the makeup gas flow dropped from 0.01 m$^3$/min to 0.004 m$^3$/min.  At the end of the purge, 203 m$^{3}$ had passed through the cryostat, corresponding to 8.2 volume exchanges.  The oxygen level was reduced to 5.2 ppm, the water concentration was reduced to 0.99 ppm, and the nitrogen concentration was reduced to 13.4 ppm.  Figure~\ref{figure:gas_argon_purge} shows the concentrations of water and oxygen during the gaseous argon purges for both run periods, along with the results from the capillary tube oxygen measurements from the first run period.  The argon purges for both run periods delivered similar results.  

\begin{figure}
\begin{center}
\includegraphics[width=1.0\linewidth]{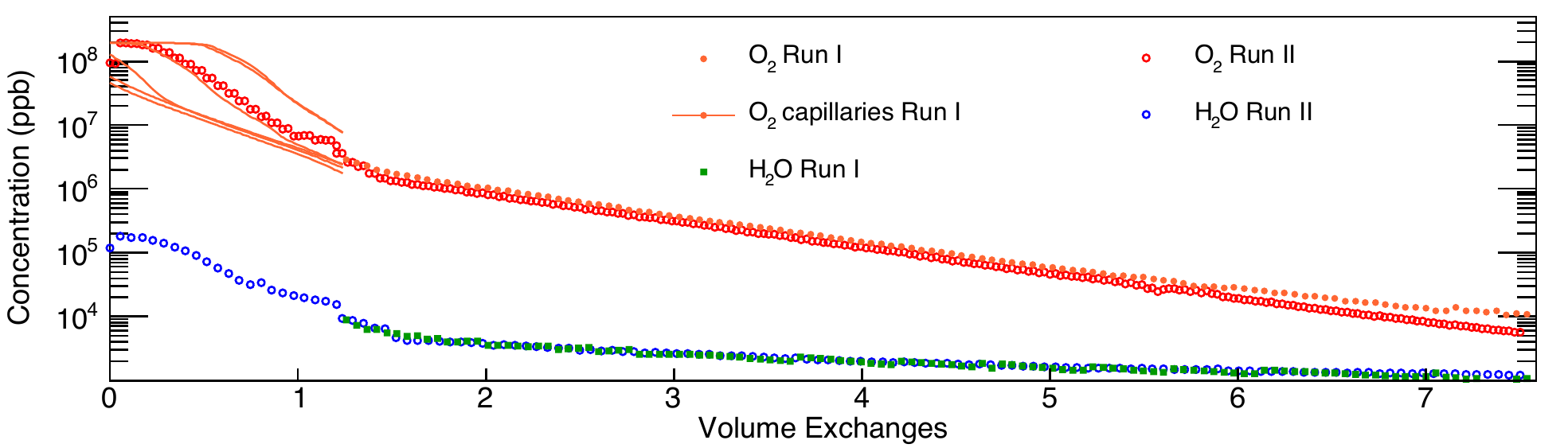}
\caption{The water and oxygen concentrations in the LAPD during the two gaseous argon purges as a function of the number of volume exchanges.  The plot shows the water concentration and the oxygen concentration measured by both the gas analyzer and the oxygen capillary tubes.  Similar results were obtained for both purges.  } 
\label{figure:gas_argon_purge}
\end{center}
\end{figure}

\subsection{Gas Recirculation}

After the removal of the ambient air from the argon purge, argon gas was pumped through the molecular sieve and oxygen filter at a rate of a volume exchange every 3.4 hours, then returned to the cryostat.  The gas recirculation for the second run period lasted for about 77 volume exchanges corresponding to one week.  Figure~\ref{figure:gas_recirculation} shows the oxygen and water concentrations, measured by the water and oxygen gas analyzers, for the gas recirculation phase.  At the end of this phase the oxygen concentration was reduced to approximately 20 ppb.  At 40 volume exchanges the oxygen outgassing rate was $2.22\times10^{-7}$~g/sec, which decreased to $5.87\times10^{-8}$~g/sec at the end of the gas recirculation phase.  The measured water concentration stabilized at about 667 ppb after 60 volume exchanges, which corresponds to an outgassing rate of $1.03\times10^{-6}$~g/sec.  The outgassing rates estimated during this phase are a lower bound.  The gas recirculation intercepted some outgassing upstream of the analyzer measurements such that the calculated rates based on analyzer measurements are likely lower than the actual rates.  Nitrogen is not appreciably removed by the filters and the nitrogen concentration at the end of this operational phase was roughly equal to the concentration achieved at the end of the purge.  The overall results for the gas recirculation phase indicate that water outgasses from all surfaces of the cryostat and piping, and that the water outgassing rate is matched by the filtration rate after several volume exchanges while the oxygen outgassing rate continues to decline.  The gas recirculation phase provides an important opportunity to debug the system and look for any final leaks before committing to cryogenic operations.

\begin{figure}
\begin{center}
\includegraphics[width=1.0\linewidth]{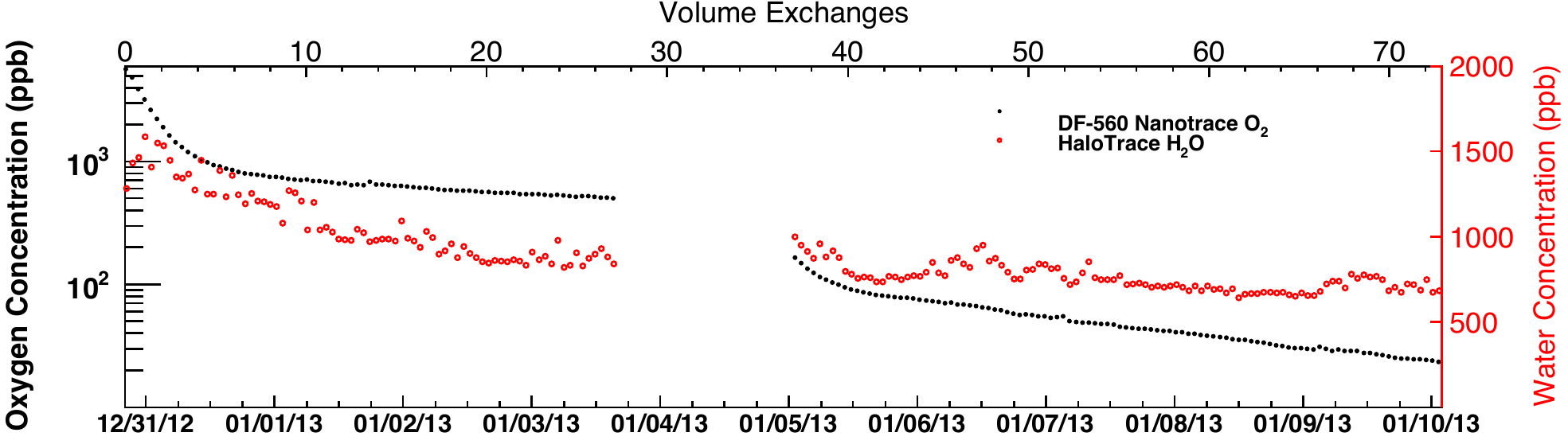}
\caption{The water and oxygen concentrations in the cryostat gas during the gas recirculation phase of the second period, plotted as a function of time and of the number of volume exchanges.  The stabilization of the oxygen contamination level at 500 ppb indicated a leak in the system that was investigated and repaired between the 27$^{th}$ and 37$^{th}$ volume exchanges.}
\label{figure:gas_recirculation}
\end{center}
\end{figure}

\subsection{Liquid Argon Filling}

For the second run period, the cryostat was filled with liquid argon from the D0 calorimeter at Fermilab. Filling the cryostat occurred in multiple stages.  The duration of each fill varied from 4 to 6 hours.  The four trailers were delivered over a period of two weeks in January, 2013.  The total volume of liquid argon placed in the cryostat was 21,309 liters, corresponding to 29.7 tons. Prior to each fill the concentrations of oxygen, water, and nitrogen were measured at the supply trailer before introducing the liquid argon into the LAPD.  The D0 liquid argon was found to contain 200 ppb oxygen and 9.5 ppm nitrogen.  The concentration of water in the D0 liquid argon was measured as less than 4 ppb which is the lower detectable limit of the gas analyzer.  A water concentration below the measurable limit of the analyzer is the typical result when sampling from the liquid phase.  The cryostat was filled through the filters such that the majority of the electronegative impurities were removed before the liquid argon reached the cryostat.



\subsection{Liquid Argon Recirculation}

After the cryostat was full, the liquid recirculation pump was started and filtration proceeded by routing liquid argon through the two filters.  Figure~\ref{figure:liquid_argon_recirc} shows the oxygen concentrations in the cryostat liquid as measured by the oxygen analyzer as a function of time for two select intervals. The first interval was after several interventions where the ullage space was exposed to atmosphere which increased the liquid argon oxygen concentration to 978 ppb.  The oxygen filter saturated which caused the deviation from the perfect mixing simulation, where perfect mixing is defined as the condition where there is no concentration gradient in the cryostat.  The second interval was after the oxygen filter was regenerated.  After oxygen filter regeneration the filter was able to quickly remove the remaining oxygen at a rate that matched the perfect mixing simulation.  

\begin{figure}
\begin{center}
\includegraphics[width=1.0\linewidth]{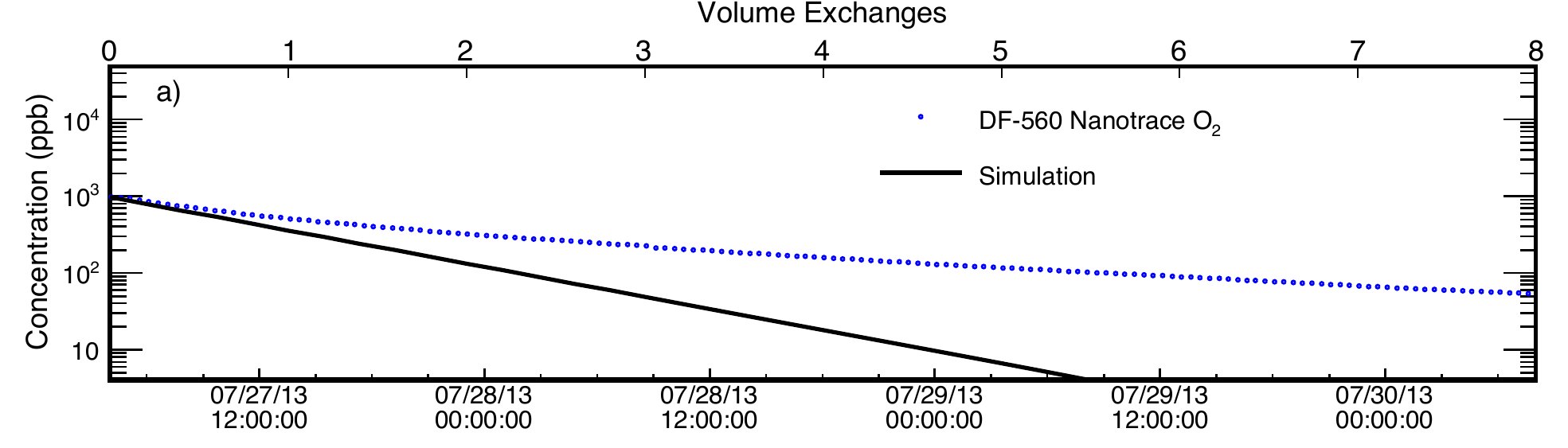}
\includegraphics[width=1.0\linewidth]{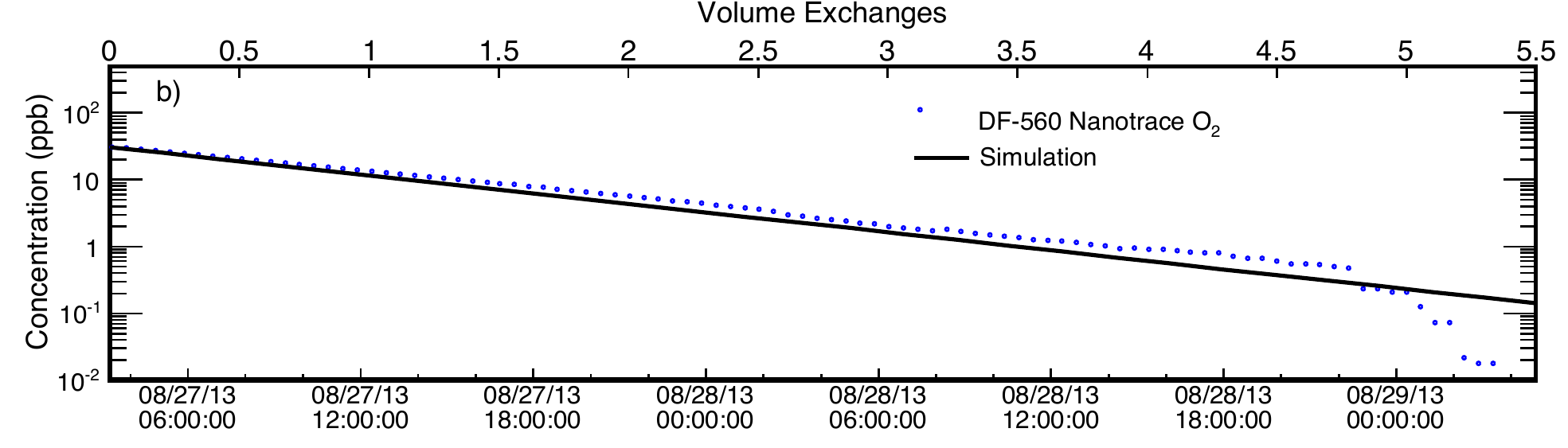}
\caption{The oxygen concentration in the cryostat liquid for two time intervals.  The measured oxygen concentration is compared to a simulation assuming perfect mixing (black line).}
\label{figure:liquid_argon_recirc}
\end{center}
\end{figure}

Figure~\ref{figure:water_oxygen_temp_gas} shows the water and oxygen concentrations measured in the cryostat vapor space with the cryostat totally filled with liquid argon, along with the temperature in the PC4 hall.  The temperature and the water concentration in the cryostat vapor space are closely correlated and suggests significant outgassing of water in this region.  This is consistent with results obtained from the MTS and suggests that the majority of the contaimination introduced into the liquid can be attributed to the contamination in the vapor space.  

\begin{figure}
\begin{center}
\includegraphics[width=1.0\linewidth]{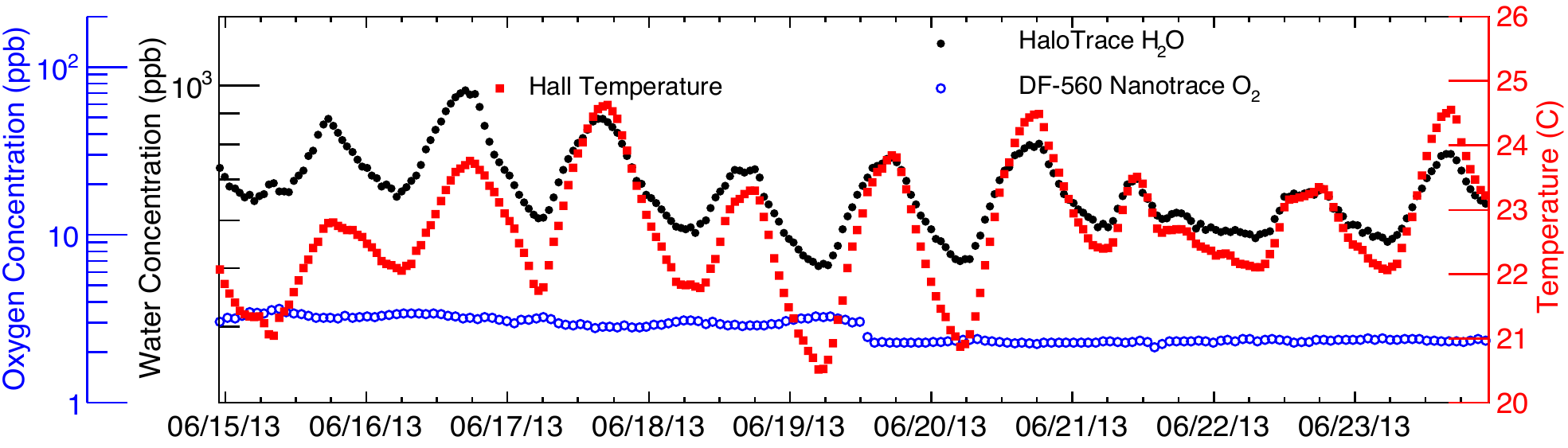}
\caption{The water concentration (solid circles) and oxygen concentration (open circles) measured in the cryostat vapor space with the cryostat totally filled with liquid argon.  Also shown is the temperature in the PC4 hall (solid squares) for the same time period. }
\label{figure:water_oxygen_temp_gas}
\end{center}
\end{figure}

After several liquid volume exchanges, the contamination was sufficiently low to begin operation of the four purity monitors inside the cryostat and the inline purity monitor upstream from the filters.  The results presented in this paper were obtained using the short purity monitor located at the periphery of the tank.  Of the five purity monitors, this one showed the best performance and provided measurements with the highest precision.  

Figure~\ref{figure:prm4_atten} shows the cathode peak pulse height, $Q_{C}$, and the ratio of the anode and cathode peak pulse height, $Q_{A}/Q_{C}$, over the complete LAPD run with several periods of extended stable running.  We measure a drift time of 0.31~ms, corresponding to the time difference between the cathode peak and anode peak.  The drift time depends on the purity monitor applied voltage and the type of purity monitor.  Using this drift time value measured by the short peripheral purity monitor, measured electron drift lifetimes of more than 6 ms were sustained for several periods of many weeks.  With the exception of the long peripheral purity monitor which showed significant and immediate photocathode degradation, the other purity monitors exhibited performance similar to that shown in Figure~\ref{figure:prm4_atten}.  The cathode purity monitor signal appears to become less efficient over time.  Nevertheless, even with low cathode signals, measured lifetime values are consistent with those at times during which good signals were achieved.  

The pump speed, and thus the volume exchange rate were changed over several time intervals throughout the run.  We found no correlation between volume exchange rate and measured lifetime values.  This behavior indicates that the boil off argon vapor intercepts the majority of the outgassing contamination from the ullage before the contamination can diffuse downward into the liquid argon.  Although the facility was not equipped to test this hypothesis, the absence of an observable effect of filtration rate on the electron lifetime indicates that filtration of the boil off gas alone could sustain the purity once the bulk liquid was filtered.  

\begin{figure}
\begin{center}
\includegraphics[width=1\linewidth]{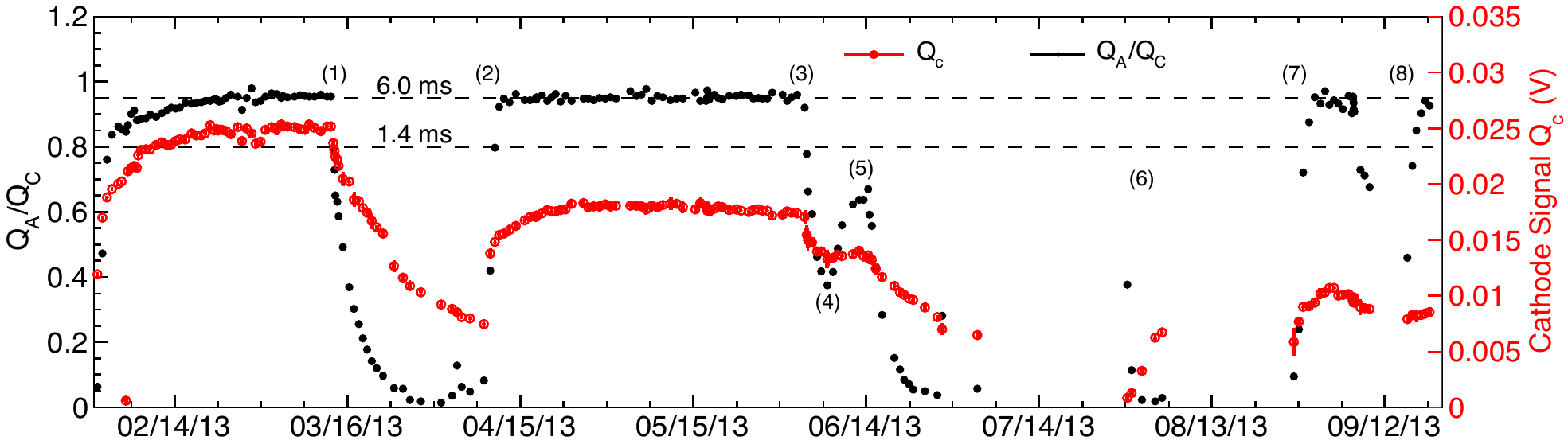}
\caption{The cathode signal ($Q_{C}$) indicated as open red circles and the anode-to-cathode ratio ($Q_{A}/Q_{C}$) indicated as solid black circles for recirculated liquid argon over all LAPD running.  The anode-to-cathode ratio is a measure of the electron lifetime, and the values $\tau=1.4$ ms and $\tau=6.0$ ms that correspond to $Q_A/Q_C=0.80$ and $Q_A/Q_C=0.95$, respectively, are indicated with dashed lines. Gaps in the data occur when either the purity monitors do not have sufficient resolving power or when they were not operating.  Special events are enumerated with the following descriptions:  (1) Circulation pump trip for an extended time period.  (2) Beginning of contamination cleanup.  (3) Pump trip lasting one hour resulting in subsequent zero flow to the filters  (4) Start of flow to filters after the pump trip. (5) Stopped pump for maintenance.  (6) Start of second cleanup, see Figure~20a.  (7) Start of third cleanup, see Figure~20b.  (8) Pump restart after a few-day period to insert a digital camera.  }
\label{figure:prm4_atten}
\end{center}
\end{figure}

\subsection{Filter Capacity}

During the second run period the oxygen filter saturated as shown in Figure~\ref{figure:liquid_argon_recirc}. Saturation was confirmed by the failure of the pumped liquid filtration to reduce the oxygen level to the zero point of the gas analyzer while sampling the cryostat liquid, identical gas analyzer readings upstream and downstream of the oxygen filter indicating lack of filtration, and the inability of the purity monitors to resolve a lifetime.  

The loading of the oxygen filter was integrated from the start of the second run period until the saturation was observed.  The contributions are shown in Table~\ref{table:loading}.  The primary contribution is what the table refers to as pumped liquid filtration, which includes the oxygen present in the cryostat liquid at the start of each period of pumped liquid filtration.  The majority of this oxygen was introduced during periods where the liquid pump was off, by opening  flanges  that  access  the cryostat ullage during insertions and extractions of hardware.  These accesses introduced significant oxygen  into  the  tank  ullage  which  subsequently  mixed into the tank liquid.   

At various intervals during periods of pumped liquid filtration the oxygen concentration in the tank vapor space was measured  near  the  point  at  which vapor exits the tank on its way to the condenser.  This oxygen concentration was roughly constant at about 8 ppb during the entire run such that the oxygen outgassing rate can be calculated and was estimated as $1.51\times10^{-7}$~g/sec.   

The capacity of the molecular sieve for water is much more difficult to determine.    All  attempts  to  measure  the  concentration  of  water  in  the  liquid argon from a capillary inserted into the liquid argon in which the liquid changes to gas which is then analyzed result in a zero reading by the commercial gas analyzers which are capable of single digit ppb water measurements.  Thus the concentration of water in the supply liquid argon or subsequently at the start of periods of pumped liquid filtration could not be determined.   

The amount of water removed from the tank during the room temperature gas recirculation is straight forward to estimate and included in Table~\ref{table:loading}.  However, because water is attracted to metallic surfaces and the large amount of metallic surface area available between the cryostat and the molecular sieve it is unknown how much of this water made it to the molecular sieve.   

Along with the oxygen concentration, the water concentration was also measured at various intervals during periods of pumped liquid filtration in the tank vapor space near the point at which vapor exits the tank on its way to the condenser.  This water concentration was roughly constant at about 14 ppb during the entire run such that the water outgassing rate can be calculated and was estimated as $1.49\times10^{-7}$~g/sec.  Argon vapor with this water concentration is condensed and then sent to the pump suction and subsequently the molecular sieve filter.  However due to the affinity of water for metallic surfaces and the extremely low vapor pressure of water at liquid argon temperatures it is unclear how much of the water in the cryostat ullage made it to the molecular sieve filter.   

It should be noted that the filters operate in series and that the molecular sieve can remove trace amounts of oxygen and conversely the oxygen filter can remove water.  Thus  the  filter  capacity  in  this  service  may  be  contingent  upon  similarly  sized filters in series.   

\begin {table*}
\centering
\caption{Capacity of the oxygen and water filters.  Each row in the table indicates the amount of contamination presented to the filters during the indicated phases of operation.  The gas recirculation phase indicates the filter loading at room temperature, all other phases are for liquid argon temperatures. It is difficult to estimate the amount of water removed from the liquid}
\begin {tabular} {lcc}
\hline
\hline
                                                                                                    & Oxygen (g) & Water (g) \\
\hline
 Gas recirculation                                                                        & 0.38          & 0.98  \\
 Cryostat filling                                                                            & 5.68           & --       \\
 Integrated ullage outgassing during pumped liquid filtration     & 1.41           & 1.39  \\
 Pumped liquid filtration                                                              & 24.09         & --       \\
 Total                                                                                           & 31.56         & 2.37  \\
 gram species per kilogram filter material                                  & 0.49           & 0.04  \\
\hline
\hline
\end {tabular}
\label{table:loading}
\end{table*}

\section{Discussion and Conclusion}\label{sec:discussion}

The tests performed using the LAPD were motivated by the need to reduce costs associated with the construction of an evacuable cryostat for future multi-kiloton detectors.  The primary goal of the LAPD has been achieved using a three stage approach including a gaseous argon purge, followed by stages of gas and liquid recirculation.  Measurements from purity monitors installed in the cryostat establish for the first time that purities that allow for electron drift lifetimes greater than 6~ms can be achieved in a large volume of liquid argon without first evacuating the vessel.  In addition to demonstrating that evacuation is not necessary for achieving long electron lifetimes, we showed that temperature gradients in the bulk volume behaved as expected when compared to FEA calculations.  We also saw that varying the liquid argon volume exchange rates did not affect the ability of the system to keep the argon free of electronegative contamination, nor does the introduction of a TPC and its associated cables.  We have also provided a measurement of the loading capacity of the filter material used in this test.



\noindent
\newline
Acknowledgments  \\

\noindent We thank the staff at Fermilab for their technical assistance in running the LAPD experiment. We acknowledge support by the Grants Agencies of the DOE.








\end{document}